\documentclass[aps,prl,twocolumn,superscriptaddress,showpacs,showkeys]{revtex4-1}

\usepackage{graphicx}
\usepackage{amsmath, amsxtra, amssymb, mathtools, isomath, nccmath, xspace, siunitx}
\usepackage{color}
\usepackage[utf8]{inputenc}
\usepackage[british]{babel}
\usepackage{hyperref}
\usepackage{sidecap}

\renewcommand{\vec}{\vectorsym}

\newcommand{\ket}[1]{\ensuremath{\lvert #1 \rangle}\xspace}%
\newcommand{\bra}[1]{\ensuremath{\langle #1 \rvert}\xspace}%
\newcommand{\alat}{\ensuremath{a_{\text{lat}}}\xspace}%
\newcommand{\gtwo}{\ensuremath{g^{(2)}}\xspace}%
\newcommand{\DelCol}{\ensuremath{\Delta_{\vec{i}}^{(\mathrm{coll})}}\xspace}
\newcommand{\matfour}[4]{\left[ \begin{array}{cc} #1 & #2 \vspace{0.2cm}\\ #3 & #4  
\end{array}\right]}
\newcommand{\br}[1]{\left(#1\right)}
\newcommand{\brs}[1]{\left[#1\right]}

\newcommand{\ii}{\mathrm{i}}
\newcommand{\ee}{\mathrm{e}}

\newcommand{\pdd}[2]{\frac{\mathrm{d} #1}{\mathrm{d} #2}}
\newcommand{\J}{\hat{\mathcal{J}}}
\newcommand{\C}{\hat{\mathcal{C}}}
\newcommand{\Jd}{\J^\dagger}
\newcommand{\Cd}{\C^\dagger}
\newcommand{\comm}[2]{\left[#1, #2\right]}
\newcommand{\expval}[1]{\langle #1 \rangle}

\newcommand{\Rydblvl}{e}

\sidecaptionvpos{figure}{h}

\begin{document}
\title{\bf{Many-body interferometry of a Rydberg-dressed spin lattice}}

\author{Johannes Zeiher}
\email[]{johannes.zeiher@mpq.mpg.de}
\affiliation{Max-Planck-Institut f\"{u}r Quantenoptik, 85748 Garching, Germany}
\author{Rick van Bijnen}%
\affiliation{Max-Planck-Institut f\"{u}r Physik komplexer Systeme, N\"{o}thnitzer Stra\ss e 38, 01187 Dresden, Germany}
\author{Peter Schauß}
\thanks{present address: Department of Physics, Princeton University, Princeton, New Jersey 08544, USA}
\author{Sebastian Hild}
\author{Jae-yoon Choi}
\affiliation{Max-Planck-Institut f\"{u}r Quantenoptik, 85748 Garching, Germany}
\author{Thomas Pohl}
\affiliation{Max-Planck-Institut f\"{u}r Physik komplexer Systeme, N\"{o}thnitzer Stra\ss e 38, 01187 Dresden, Germany}
\author{Immanuel Bloch}%
\affiliation{Max-Planck-Institut f\"{u}r Quantenoptik, 85748 Garching, Germany}
\affiliation{Ludwig-Maximilians-Universit\"{a}t, Fakult\"{a}t f\"{u}r Physik, 80799 M\"{u}nchen, Germany}%

\author{Christian Gross}%
\affiliation{Max-Planck-Institut f\"{u}r Quantenoptik, 85748 Garching, Germany}

\date{\today}


\begin{abstract}
Ultracold atoms are an ideal platform to study strongly correlated phases of matter in and out of equilibrium.
Much of the experimental progress in this field crucially relies on the control of the contact interaction between two atoms.
Control of strong long-range interactions between distant ground state atoms has remained a long standing goal, opening the path towards the study of fundamentally new quantum many-body systems \cite{Baranov2012, Balents2010} including frustrated or topological magnetic phases~\cite{Glaetzle2015,vanBijnen2015} and supersolids \cite{Henkel2010, Pupillo2010}.
Optical dressing of ground state atoms by off-resonant laser coupling to Rydberg states has been proposed as a versatile method to tailor such interactions \cite{Santos2000, Bouchoule2002, Henkel2010, Pupillo2010}.
However, up to now the great potential of this approach for interaction control in a many-body setting has eluded experimental confirmation.
Here we report the realisation of coherent Rydberg-dressing in an ultracold atomic lattice gas and directly probe the induced interaction potential using interferometry combined with single atom sensitive detection.
We implement a two-dimensional synthetic spin lattice of near unity filling and demonstrate its versatility by tuning the range and anisotropy of the effective spin interactions.
Our measurements are in remarkable agreement with exact solutions of the many-body dynamics, providing further evidence for the high degree of accurate interaction control in these systems.
Our work marks the first step towards the use of laser-controlled Rydberg interactions for the study of exotic quantum magnets\cite{Glaetzle2014,Glaetzle2015,vanBijnen2015} in optical lattices.
\end{abstract}
\maketitle


Neutral ultracold atoms in optical lattices are among the most promising platforms for the implementation of analog quantum simulators of condensed matter systems.
However, the simulation of magnetic Hamiltonians, often emerging as an effective model in more complex many-body systems, is difficult with contact interactions due to the low energy scale of the associated superexchange process~\cite{Trotzky2008}.
Long-range interactions offer an alternative way to directly achieve strong effective spin-spin interactions.
Such interactions emerge between magnetic atoms and between ultracold polar molecules~\cite{Lahaye2009}, trapped ions~\cite{Blatt2012} or ground state atoms resonantly~\cite{Saffman2010} or off-resonantly coupled (``dressed'') to Rydberg states~\cite{Santos2000,Bouchoule2002,Henkel2010,Pupillo2010}.
Rydberg dressing is especially appealing due to the simplicity of realising atomic lattice systems with unity filling combined with the great tunability of the interaction strength and shape, which might be exploited to explore exotic models of quantum magnetism~\cite{Glaetzle2015, vanBijnen2015}.
While effects of long-range spin interactions have been observed in many-body systems of polar molecules~\cite{Yan2013}, ions ~\cite{Britton2012a,Richerme2014a,Jurcevic2014a} and resonantly excited Rydberg atoms~\cite{Schauss2015, Labuhn2016}, neither of these approaches combines the advantages of Rydberg dressing, which permits to realise strong spin interactions in lattices with near unity filling. So far, Rydberg dressing in a many-body system remains an experimental challenge, for which up to now only dissipative effects have been measured~\cite{Weber2012,Balewski2014,Malossi2014,Schempp2014,Goldschmidt2015}.
For two atoms, first promising experimental results have been reported recently for near-resonant strong dressing~\cite{Jau2016}, where, however, the assumption of weak Rydberg state admixture required for the realisation of various many-body models~\cite{Henkel2010,Pupillo2010,Mattioli2013,Glaetzle2014,Li2015a} does not hold~\cite{Honer2010}.

Here we demonstrate Rydberg-dressing in a two-dimensional (2d) near-unity filled atomic lattice with tailored extended range interactions between approximately $200$ effective spins.
In contrast to our previous experiments\cite{Schauss2012,Schauss2015} on resonantly coupled Rydberg gases, all atoms participate here in the spin dynamics. 
We exploit the temporal control over such interactions to perform interferometric measurements and directly reveal the induced correlations via single-site sensitive local detection. 
Our experiments illustrate the versatility of Rydberg-dressing by tuning the range and isotropy ~\cite{Carroll2004,Barredo2014} of the interaction potential induced by optical coupling to high-lying $P$-states. 
First-principle potential calculations together with an exact solution of the many-body dynamics accurately reproduce these measurements. 
From this we conclude that stimulated Rydberg-Rydberg state transitions, recently observed as anomalous broadening~\cite{Goldschmidt2015}, act collectively in a many-body system, implying that rare one-body decay events can globally affect the entire spin lattice.


\begin{figure*}
  \centering
  \includegraphics[width=\textwidth]{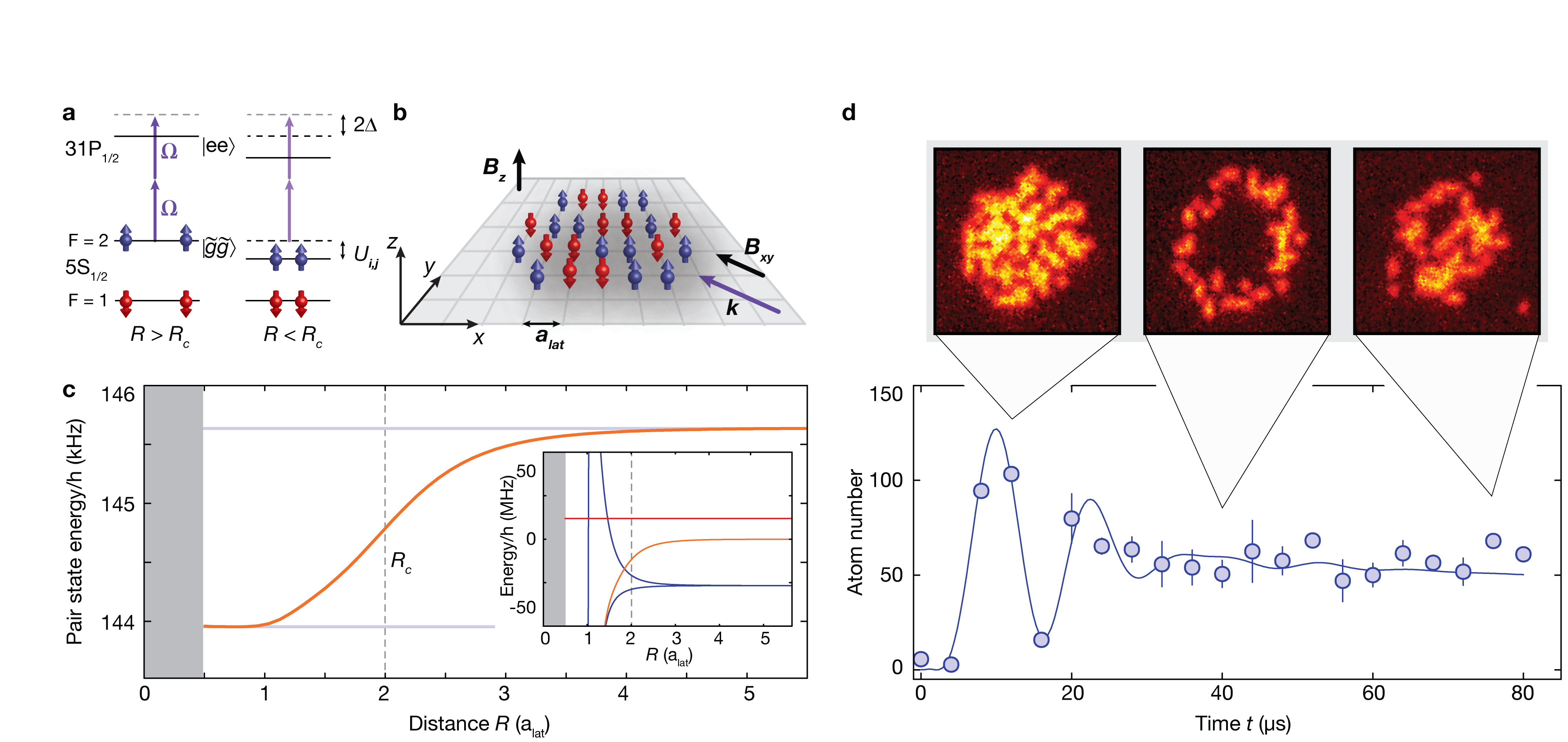}%
  \caption{ \label{fig:1}
  \textbf{Schematic of the experiment.}
  \textbf{a}, The $5S_{1/2}$ $\ket{F=2,m_F=-2}$ state (spin up $\ket{\uparrow}$, blue arrow) is coupled to a Rydberg state $\ket{e}$ in the manifold of $31P_{1/2}$ with Rabi frequency $\Omega$ (purple arrow) and detuning $\Delta$ (upper black arrow), leading to the dressed state $\ket{\tilde{g}}$.
For separations $R<R_c$, the bare Rydberg interactions detune the state $\ket{ee}$ with two atoms in $\ket{e}$ (shifted black solid line), which is two-photon coupled via the singly excited intermediate state.
 This induces interactions $U_{\vec{i},\vec{j}}$ between ground state atoms in $\ket{\uparrow}$ whereas the spin down state $\ket{\downarrow}$ ($\ket{1,-1}$, red arrow) remains unaffected.
    \textbf{b}, The excitation laser propagates along $\vec{k}$ in the plane of the 2d spin lattice (lattice constant $\alat$) at an angle of $45^{\circ}$ with respect to the $x$ and $y$ axes.
  The magnetic field $\vec{B_z}$ ($\vec{B_{xy}}$) was aligned with the $z$-axis ($\vec{k}$).
   \textbf{c}, Calculated soft-core potential for $31P_{1/2}$ for $\Omega_s/2\pi=1.33\,\mathrm{MHz}$ and $\Delta/2\pi=6\,\mathrm{MHz}$ (orange solid line).
The AC-Stark shift of the pair state and the soft-core saturation value $U_0$ are shown as the upper and lower horizontal light blue line respectively.
 The inset shows the bare Rydberg interaction curves (blue solid lines) including the optically coupled potential (orange solid line) underlying the soft-core interaction.
  $R_c$ (vertical dashed line) marks the distance where the interaction shift equals $2\Delta$ (red solid line). The grey shading indicates the region where crossings of other Rydberg states are expected, leading to slight modifications of the soft-core at distances well below $\,\alat$ (Fig.~\ref{fig:S4}).
 \textbf{d}, Ramsey oscillation fringe with $\ket{\uparrow}$ coupled to $31P_{3/2}$, $m_J=-3/2$ with $\Omega_s/2\pi=1.9(1)\,\mathrm{MHz}$ and $\Delta/2\pi=-8\,\mathrm{MHz}$ (blue points).
 Three representative single shots of the atom distributions at the times indicated by the grey triangles are shown above.
 In the central shot, the bulk part of the sample had acquired a relative phase shift of $\pi$ with respect to particles situated at the edge due to the collective longitudinal field $\DelCol$.
 The theory prediction (solid blue line) reproduces the interaction induced dephasing of the oscillation.
 The errorbars denote the standard error of the mean (s.e.m).}
\end{figure*}

Dressed atomic states emerge as new eigenstates of an atom driven by a laser field with a  Rabi frequency $\Omega$ and frequency detuning $\Delta$. Considering two levels $\ket{g}$ and $\ket{e}$ in the weak dressing regime, $\Omega \ll \Delta$, the dressed ground state $\ket{\widetilde{g}}\approx\ket{g}+\beta\ket{e}$ contains a small admixture $\beta=\frac{\Omega}{2\Delta}$ of the state $\ket{e}$. As a result, it acquires a finite lifetime $\tau_r/\beta^2$, which, however, greatly exceeds the lifetime $\tau_r$ of the bare excited state.
Choosing $\ket{e}$ to be a Rydberg state, the laser-coupling also induces effective interactions between two dressed atoms in state $\ket{\widetilde g}$ (Fig.~\ref{fig:1}a). At large interatomic distances $\vec{R}$ this interaction, $U(\vec{R})\approx\beta^4 V(\vec{R})$, corresponds to the Rydberg-Rydberg atom interaction potential $V(\vec{R})$ reduced by the probability $\beta^4$ to excite both atoms at once. At short distances, however, the strong interaction between Rydberg atoms blocks this simultaneous excitation within a critical distance, $R_c$, determined by $V(R_c)=2\hbar\Delta$. As a result, the induced interaction acquires a soft-core shape and saturates to a value of $U_0=\hbar\Omega^4/(8|\Delta|^3)$ (Fig. \ref{fig:1}c)~\cite{Bouchoule2002,Henkel2010,Pupillo2010}.
Involving other atomic ground states in the dynamics, one naturally obtains various kinds of lattice models of interacting spins that have been proposed for metrology applications~\cite{Bouchoule2002, Gil2014} or the exploration of exotic quantum magnetism~\cite{Glaetzle2015, vanBijnen2015}. In the simplest case, a single additional ground state that is not coupled to the Rydberg state (Fig. \ref{fig:1}a, b) results in a system described by a 2d Ising Hamiltonian
\begin{eqnarray}
\hat{H}=\hbar\sum_{\vec{i}}^{N}\left(\delta + \DelCol\right)\hat{S}^{z}_{\vec{i}}+\sum_{(\vec{i}\neq \vec{j})}^N \frac{U_{\vec{i},\vec{j}}}{2}\hat{S}^{z}_{\vec{i}}\hat{S}^{z}_{\vec{j}}. \label{Eq1}
\end{eqnarray}
Here, $\hat{S}^{z}$ is a spin-$1/2$ operator with eigenstates $\ket{\uparrow}$ and $\ket{\downarrow}$, corresponding to the Rydberg-dressed and uncoupled atomic ground state, respectively.
The longitudinal field $\delta$ arises from the single atom light shift, $\delta\approx \Omega^2/(4\Delta)$ and $U_{\vec{i},\vec{j}}=U_0/(1+(R/R_c)^6)$ denotes the dressing-induced interaction between spins located on lattice sites $\vec{i}$ and $\vec{j}$ at a distance  $R/\alat=|\vec{i}-\vec{j}|$, where \alat denotes the lattice constant. The collective contribution $\DelCol=\sum_{\vec{j}\neq \vec{i}}^N \frac{U_{\vec{i},\vec{j}}}{2}$ results from the transformation from the original atomic-state representation to effective spin operators~\cite{Schachenmayer2010, Schauss2015}. The interaction induced energy shift resulting from this additional longitudinal field is measurable as a frequency shift in an interferometric Ramsey sequence (Fig.~\ref{fig:S1},\ref{fig:S2}). Furthermore, due to the extended range of the interactions, this collective field depends on the spins nearby and hence spins near the edge of a finite system evolve differently compared to spins in the center, which we clearly observe (Fig. \ref{fig:1}d).
The optically switchable spin interactions $U_{\vec{i},\vec{j}}\propto \Omega(t)^4$ enable such interferometric measurements of the many-body system by sequential application of the interaction Hamiltonian~\eqref{Eq1} and a transverse magnetic field induced by microwave coupling of the two ground states~\cite{Gil2014, Mukherjee2015}.

\begin{figure}
  \centering
  \includegraphics[width=0.45\textwidth]{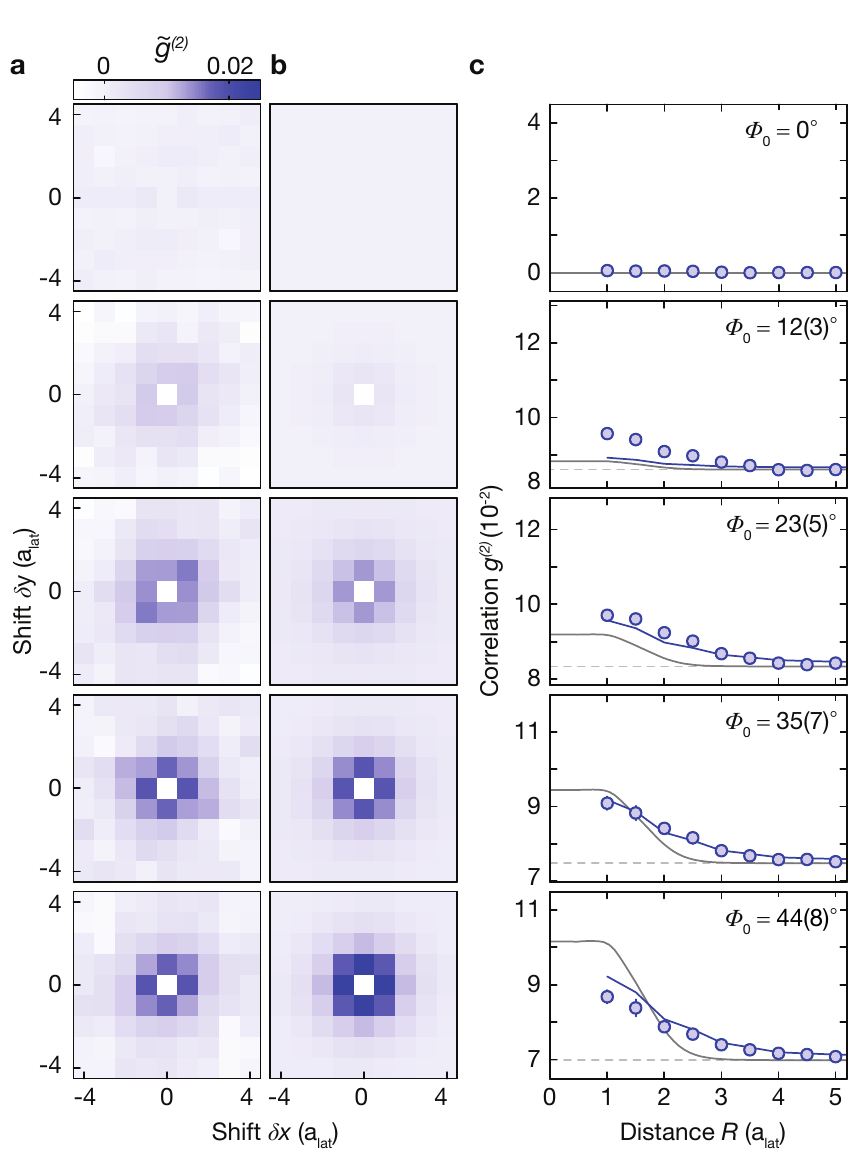}%
  \caption{ \label{fig:2} 
  \textbf{Interferometric measurement of spin-spin correlation.}
  \textbf{a}, Measured 2d spin correlation functions $\tilde{g}^{(2)}(\vec{R})=\gtwo(\vec{R})-\gtwo_{\infty}$ for increasing interaction phase $\Phi_0$(top to bottom) evaluated in a region of interest of $9\times9$ lattice sites. The constant spatial offset $\gtwo_{\infty}$ was obtained by azimuthally averaging $\gtwo(\vec{R})$ for $4.5\leq|\vec{R}|/\alat\leq7.5$ and subtracted for each dataset.
  The two axes are the shifts $\vec{R}=(\delta x,\delta y)$.
  \textbf{b}, Theory prediction for $\tilde{g}^{(2)}(\vec{R})$ including dissipation with $\gtwo_{\infty}$ adjusted to the experimentally determined value.
  \textbf{c}, Azimuthal averages of the 2d correlations in a (blue points), and b (blue solid line) versus distance $R=|\vec{R}|$.
  The perturbative estimate (solid dark grey line) shifted by $\gtwo_{\infty}$ (dashed light grey line) differs significantly from the non-perturbative prediction for the spin correlation (see main text). The shown correlations were extracted from $250-300$ experimental shots.
  All errorbars denote the s.e.m..}
\end{figure}

Our experiments started with a 2d degenerate gas of rubidium-$87$ in the $\ket{F,\,m_F}=\ket{1,-1}$ hyperfine state, confined in a single antinode of a vertical ($z\,$-axis) optical lattice.
In this single $x-y$ plane, we then switched on a square optical lattice with $\alat=532\,$nm spacing and prepared about $190$ atoms in a unity filling Mott insulator with a defect fraction of about $3\%$.
The chosen atom number ensured a negligibly small number of doubly occupied sites.
Transitions from the state $\ket{1, -1}$ (``spin down'', $\ket{\downarrow}$) to $\ket{2, -2}$ (``spin up'', $\ket{\uparrow}$) were driven globally via microwave pulses.
In order to introduce long-range interactions, the state $\ket{\uparrow}$ was laser coupled to the $31P_J$ state ($J=1/2$ or $J=3/2$), which has a lifetime of approximately $\tau_r=27\,\mu$s (see ref. \cite{Beterov2009}).
The excitation beam at a wavelength of $297\,\mathrm{nm}$ propagated in the plane of the atoms along the diagonal of the cubic lattice (Fig. \ref{fig:1}b).
A static magnetic field was used to set the quantisation axis either along the $z$-direction or aligned with the laser beam wavevector $\mathbf{k}$, which allowed for the selective coupling to different Rydberg states depending on the polarisation of the excitation beam (Fig.~\ref{fig:S3}) and Supplementary Information).
The positions of all the atoms were then detected with single lattice site resolution and single atom sensitivity~\cite{Fukuhara2013}. By optically removing atoms in state $\ket{\uparrow}$ prior to imaging, we could also perform spin-resolved detection and, in particular, measurements of spin-spin correlations.

In a first experiment, we aim to reveal the characteristic spin correlations in the many-body system that emerge over time as a result of the long-ranged spin interactions. To this end, we employ a spin-echo sequence (Fig.~\ref{fig:S1}), which is sensitive to interaction induced phase shifts, whereas the influence of single particle effects including the collective longitudinal field (Fig.~\ref{fig:1}d) is suppressed.
Starting with all atoms in state $\ket{\downarrow}$ we first applied a $\pi/2$ microwave-pulse on the $\ket{\downarrow}-\ket{\uparrow}$ transition to generate an equal superposition of the two spin states. 
This was followed by two identical Rydberg-dressing pulses of duration $t/2$ each, separated by a $\pi$ microwave-pulse. After closing the interferometer with another $\pi/2$ pulse, in the absence of interactions the system returns to its initial state with all population in $\ket{\downarrow}$, such that any deviations from this are expected to provide a precise probe of the interaction induced dynamics described by the Hamiltonian~\eqref{Eq1}. 
Optical dressing to the $31P_{1/2}$ Rydberg state was performed with a detuning $\Delta/2\pi=6\,\mathrm{MHz}$ and a Rabi frequency of $\Omega_s/2\pi=1.33(7)\,$MHz, determined independently by Ramsey spectroscopy (Fig.~\ref{fig:S2} and Supplementary Information).
These laser parameters ensure to be in the weak dressing regime with a small Rydberg state admixture of $\beta=0.11(1)$ corresponding to a population probability of $\beta^2=0.012(2)$.
The resulting long-range spin interactions with $U_0/2\pi\approx1.8\,\mathrm{kHz}$ induce correlated phase rotations during the two dressing stages peaking at $\Phi_0=\int_0^{t}U_0(\tilde{t})\,\mathrm{d}\tilde{t}$, and ultimately lead to measurable spin correlations at the end of the echo sequence. The time dependence of $U_0$ arises from the finite rise time of $\Omega$.
Our spin-resolved detection scheme provides direct access to longitudinal correlations $g_{\vec{i},\vec{j}}^{(2)}=\langle \hat{\sigma}_{\downarrow\downarrow}^{(\vec{i})} \hat{\sigma}_{\downarrow\downarrow}^{(\vec{j})}\rangle - \langle \hat{\sigma}_{\downarrow\downarrow}^{(\vec{i})}\rangle \langle \hat{\sigma}_{\downarrow\downarrow}^{(\vec{j})}\rangle$, where $\hat{\sigma}_{\downarrow\downarrow}^{(\vec{i})}$ measures the $\ket{\downarrow}$-population at site $\vec{i}$. Measurements versus time or, equivalently, versus the interaction phase $\Phi_0$ permit to trace the dynamical growth of the spin-spin correlations in the regime of small interaction phase, whereas for large $\Phi_0$ the correlation signal is expected to decrease.
Fig. \ref{fig:2}a shows the measured spin correlation function $\gtwo(\vec{R})=\sum_{\vec{i}\neq\vec{j}}\delta_{\vec{i}\vec{j},\vec{R}}g_{\vec{i},\vec{j}}^{(2)}/\sum_{\vec{i}\neq\vec{j}}\delta_{\vec{i}\vec{j},\vec{R}}$, which is obtained from a translational average constrained to a given distance $\vec{R}$ by the Kronecker symbol $\delta_{\vec{i}\vec{j},\vec{R}}$. The observed correlation functions resemble the soft-core shape of the calculated potential shown in Fig.~\ref{fig:1}c. This behaviour is readily understood in the small time limit for $N_{\mathrm{eff}}\Phi_0^2\ll1$, where $N_{\mathrm{eff}}$ denotes the number of spins within the extent $R_c$ of the interaction potential. Here, one obtains a direct proportionality $g^{(2)}(R)=\Phi_0^2/(4(1+(R/R_c)^6)^2)$ between the induced correlations and the square of the  soft-core potential.
However, due to the high filling of the system, the azimuthally averaged correlations increase beyond the value expected from this simple expression (Fig.~\ref{fig:2}c) due to the simultaneous interaction of $N_{\mathrm{eff}}\approx11$ spins, well reproduced by the exact analytical solution of the quantum dynamics (Fig.~\ref{fig:S6} and Supplementary Information). This directly highlights influence of many-body dynamics present in our system despite the seemingly small interaction phases.  

\begin{figure}
  \centering
  \includegraphics[width=0.45\textwidth]{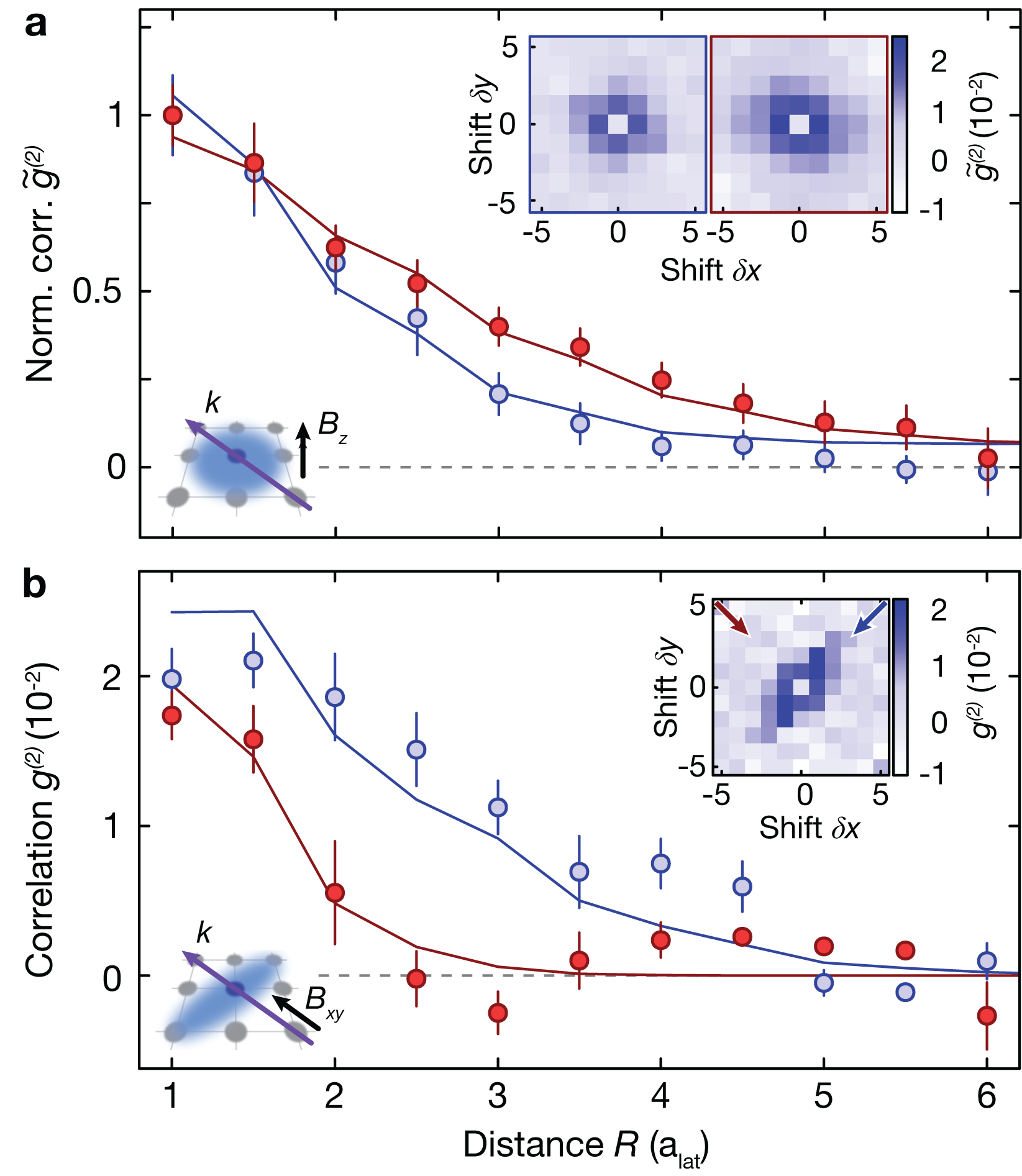}%
  \caption{\label{fig:3} 
  \textbf{Tunability of the long-range interaction.} 
  \textbf{a}, Comparison of the azimuthally averaged normalised spin correlation $\tilde{g}^{(2)}$ for $31P_{1/2}$ (blue datapoints) and $31P_{3/2}$ (red datapoints) together with theory predictions (blue and red solid lines, $\Delta/2\pi= 6\,\mathrm{MHz}$, $\Omega_s/2\pi=1.33(7)\,\mathrm{MHz}$, $\Phi_0=35(7)\,\mathrm{^{\circ}}$  ($\Delta/2\pi=-6\,\mathrm{MHz}$, $\Omega_s/2\pi=1.16(6)\,\mathrm{MHz}$, $\Phi_0=22(5)\,\mathrm{^{\circ}}$) for $J=1/2$ ($3/2$)).
  The inset shows the 2d correlations for the two cases $J=1/2$ (left) and $J=3/2$ (right).
  The inset in the lower left corner illustrates the excitation geometry with magnetic field along $z$ (black arrow) and excitation beam (purple arrow) in the atomic plane (grey discs), which leads to isotropic interaction (light blue area).
  The shown correlation was extracted from $180$ experimental shots. 
 \textbf{b}, One-dimensional averages of an anisotropic spin correlation along two orthogonal directions (red and blue datapoints), obtained by averaging two neighbouring sites, together with the prediction of the coherent theory without dissipation (red and blue solid lines, $\Omega_s/2\pi= 2.45(12)\,\mathrm{MHz}$, $\Delta/2\pi=-12\,\mathrm{MHz}$ and $\Phi_0=42(9)\,\mathrm{^{\circ}}$).
 The inset shows the 2d correlation with the two averaging directions indicated by red and blue arrows.
 Post-selection was applied before calculating the correlation on a set of $20$ images (Fig.~\ref{fig:S7}).
The inset in the lower left corner indicates the geometry as in a.
All errorbars denote the s.e.m..}
\end{figure}
Next to the optical switchability, Rydberg-dressing also enables to design the extent and anisotropy of the interactions between the spins. In the following, we demonstrate this control capability by selecting different coupled Rydberg states. As shown in Fig.~\ref{fig:3}a, changing the Rydberg state from $31P_{1/2}$ to $31P_{3/2}$ leads to notable modifications of the measured correlation function and, hence, the underlying spin interactions. Contrary to $31P_{1/2}$, the $31P_{3/2}$ state features repulsive interactions with a tenfold larger magnitude (Fig.~\ref{fig:S4}) which implies a $50\%$ larger interaction radius $R_c/\alat\approx3$ that is reflected in the enhanced correlation range shown in Fig.~\ref{fig:3}a. In both cases, the angular symmetry of the induced interactions is dictated by the strong applied magnetic field which causes isotropic interactions to emerge for the $\vec{B}_z$-configuration (Fig.~\ref{fig:1}b) used in the measurements of Fig.~\ref{fig:2}. Rotating the magnetic field permits to tailor the anisotropy of the dressing-induced interactions, which is maximised by aligning the magnetic field with the wave vector $\vec{k}$ of the circularly-polarised dressing laser in the plane of the spin lattice (Fig.~\ref{fig:S3}). The correlation measurements confirm the expected anisotropy and we observe an aspect ratio of $\sim3/2$ in quantitative agreement with the theory (Fig.~\ref{fig:3}b). 

\begin{figure}
  \centering
  \includegraphics[width=0.45\textwidth]{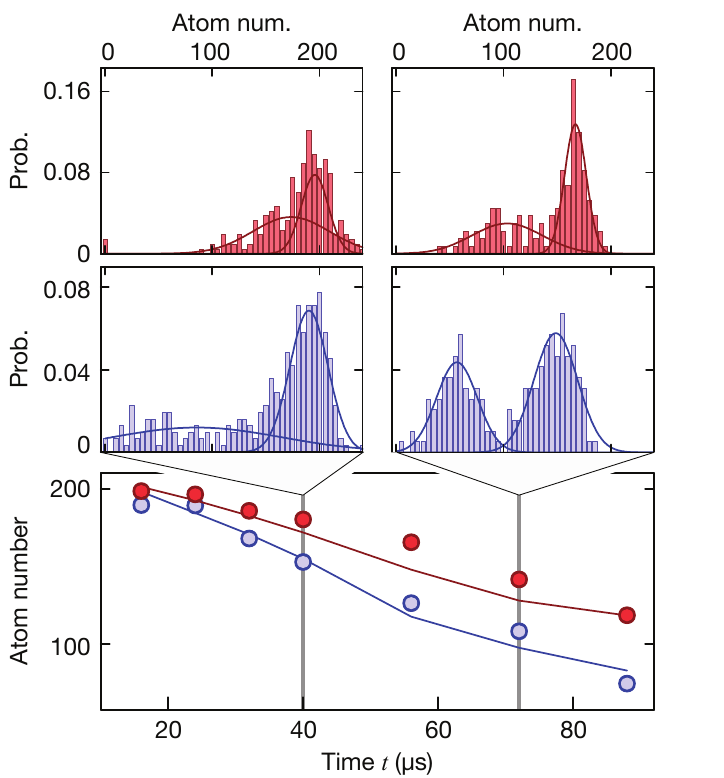}%
  \caption{\label{fig:4}
  \textbf{Collective dissipation.} 
Extracted mean atom number after a spin echo sequence versus dressing time $t$ with and without spin-resolved detection (blue and red datapoints respectively) with the theory prediction (blue and red solid lines) and corresponding number histograms above (blue and red filling) at two different times (grey vertical lines). 
  	The solid lines are Gaussians obtained by fitting their sum to the histograms.  
  All errorbars denote the s.e.m..}
\end{figure}
	
In addition to probing the coherent spin interactions, the applied spectroscopy technique allows for an in depth characterization and understanding of decoherence in the present system, which is of great importance for future applications of Rydberg-dressing.
The measured spin correlations (Fig.~\ref{fig:2}c) reveal an unexpected homogeneous offset, in addition to the characteristic spatial dependence. This requires a process affecting the system globally and can be explained in terms of an additional dissipation channel triggering the loss of all particles in the dressed $\ket{\uparrow}$ state. Such a loss process is also consistent with the measured atom number distributions that steadily develop a bimodal structure with increasing dressing time (Fig. \ref{fig:4}). Incoherent transitions to other Rydberg states due to black-body radiation present a plausible mechanism \cite{Goldschmidt2015}, since such transitions project the $\ket{\uparrow}$ state onto a nearby Rydberg state and thereby produce a real Rydberg excitation (Fig.~\ref{fig:S8}). The production of such atoms with opposite-parity with a small rate $\beta^2\gamma_{BB}$ can induce strong dipolar exchange interactions and thereby trigger a fast avalanche-like atom loss due to strong resonance broadening~\cite{Anderson1998,Mourachko1998,Goldschmidt2015}. We can incorporate this picture into our theoretical description by assuming a stochastically triggered instantaneous loss of all $\ket{\uparrow}$-state atoms as a simplification, still permitting an analytical solution of the many-body dynamics (Supplementary Information). The shape and magnitude of the measured correlations in Fig.~\ref{fig:2} is well captured by this model, while the offset is reproduced to within $20\%$.
We extract a value of $\gamma_{\rm BB}/2\pi=1.6\,{\rm kHz}$, approximately half the literature value for $31P_{1/2}$-states \cite{Beterov2009}.
Furthermore, we observe a decreasing atom number $N$ with increasing total dressing time $t$ (Fig.~\ref{fig:4}), consistent with the predicted exponential loss $N(t)=N(0){\rm e}^{-\frac{N(0)}{4}\beta^2\gamma_{\rm bb}t}$ (Supplementary Information).
Due to the dependence on system size via $N(0)$, the extracted value of $130(20)\,\mu$s is significantly lower than the anticipated value of $\tau_r/\beta^2=2.2\,\mathrm{ms}$ in the absence of additional loss processes, however it surpasses the bare Rydberg state lifetime by a factor of five.
The spin-resolved measurement agrees equally well with the predicted dynamics, in strong support for the developed understanding of the many-body dynamics.

%
In conclusion, we have implemented and detected long-range spin interactions induced by Rydberg-dressing, realising and probing a synthetic system of $\sim200$ spins.
We have demonstrated the versatility of this approach by generating attractive as well as repulsive interactions with tailored angular anisotropies between ground state atoms. The accurate control and detailed understanding of the induced interactions has been verified by comparisons to theoretical predictions obtained from first-principle calculations.
Coherence is limited by a collective decay process, which we incorporate in a model that consistently explains the observations and quantitatively reproduces all of our measurements.
We anticipate that significant improvement in coherence time is possible in designed lattice systems with fewer spins or reduced dimensionality~\cite{Schauss2015}, an appropriately adjusted Rydberg-state detuning to avoid coupling to broadened resonances, or via stroboscopic dressing that allows detrimental impurity Rydberg atoms to decay or be laser-quenched~\cite{Schauss2012} before triggering the avalanche loss.
Our results pave the way towards experimental explorations of more complex quantum magnets~\cite{Glaetzle2014,Glaetzle2015,vanBijnen2015} and the study of novel phenomena in Rydberg dressed atomic lattices \cite{Mattioli2013}.

\begin{acknowledgements}
We thank Alexander Gl\"atzle, Peter Zoller, Marc Cheneau and Nils Henkel for discussions. We acknowledge funding by MPG, EU (UQUAM, SIQS, RYSQ, Marie Curie Fellowship to J.C.) and the K\"orber Foundation.\\
\end{acknowledgements}

\bibliography{RydbergDressing_revtex_v3}

\section*{Supplementary Information}
\section{Introduction}\label{SecIntro}
This Supplementary Information document contains background information on the experimental methods and additional results followed by a discussion of the interaction potential calculation, for both undressed and dressed potentials. Towards the end, the analytical solution of the many-body dynamics is presented, with detailed derivations on how to include dissipation caused by spontaneous emission and the black-body induced decay. The theoretical curves referred to in the main text and the figures are all calculated from the analytical solution including all dissipation effects, unless stated otherwise
\section{Laser system}\label{SecLaserSyst}
\begin{figure*}
  \centering
  \includegraphics[width=0.7\textwidth]{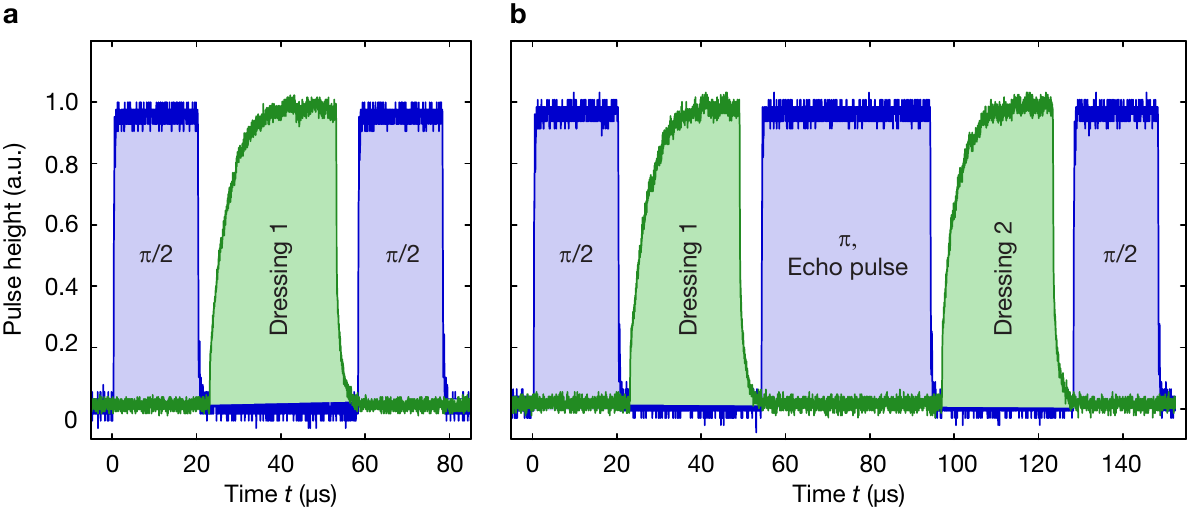}%
  \caption{\textbf{Ramsey sequences with and without spin echo.}
   \textbf{a}, Microwave pulses on the $\ket{F=1, m_F=-1}$ to $\ket{F=2, m_F=-2}$ transition with length $t_p$ and areas $\Omega_{mw}t_p=\pi/2$ are shown in blue, normalised dressing laser pulses recorded by a photo diode are shown in green. This sequence is used to measure the collective field $\DelCol$, for Fig. 1d and Fig.~\ref{fig:S2}.
   \textbf{b}, Sequence including an intermediate echo pulse with area $\pi$, used to obtain correlations shown in Figs. 2 and 3.  
  }
  \label{fig:S1} 
\end{figure*}
Excitation and coupling to the Rydberg state at $297\,\mathrm{nm}$ with a single photon has several advantages over the more conventional indirect coupling in an off-resonant two-photon configuration. 
Next to the higher achievable coupling strength and the flexibility when working with Rydberg p-states, both the light shift and the off-resonant scattering on the $D2$ line in Rubidium are negligible for direct excitation.
The required uv-light at $297\,$nm was generated in two doubling steps. We started with a diode laser at $1190\,$nm which was stabilised to a passively stable reference cavity with a finesse of approximately $10,000$.
Frequency tunability was provided by stabilising a sideband of a fibre coupled electro optic modulator on the cavity resonance. 
The light at $1190\,$nm was amplified with a Raman fibre amplifier system and frequency doubled in a single pass through a periodically poled lithium niobate crystal, providing a stable output power of up to $2\,$W at $595\,$nm.
This light was used to seed a home-built resonant doubling cavity to $297\,$nm, yielding $250-300\,$mW of optical power in the ultra violet. After spatial filtering with a $25\,\mu$m pin-hole, the light was focused on the atoms with a waist of $44(5)\,\mu$m. We estimate the power at the position of the atoms to be $45(10)\,\mathrm{mW}$.
The Rydberg resonance was measured by detecting the ground state atom loss versus laser detuning for a fixed excitation pulse time $<10\,\mu$s. Adjusting the parameters allowed us to measure resonance widths of approximately $70\,$kHz, (Fig.~\ref{fig:S2}), which is likely still limited by power broadening but sets a conservative upper bound of  $60\,$kHz on the laser linewidth at $297\,$nm.

\section{Experimental sequence and dressing pulses}\label{SecRamsey}

To probe the Rydberg-dressed interactions, we employed Ramsey interferometry. Typical sequences are shown in Fig.~\ref{fig:S1}. We prepared the system in $\ket{F=1, m_F=-1}=\ket{\downarrow}$, which was coupled to $\ket{F=2, m_F=-2}=\ket{\uparrow}$ with a microwave Rabi coupling of $\Omega_{mw}/2\pi=12.5\,\mathrm{kHz}$. 
The interferometric sequence was initialized with a microwave pulse of area $\pi/2$, preparing an equal superposition of the two spin states. Next, the dressing laser was switched on for a variable time $t/2$, before a microwave spin echo pulse of area $\pi$ was applied.
After a second dressing phase of identical duration, a final $\pi/2$ microwave pulse closed the interferometer and the $\ket{\downarrow}$ component was detected with single site resolution after a resonant pushout on the atoms in the state $\ket{\uparrow}$.
To compare our results with the theoretically expected signal, the finite rise time of the dressing pulses had to be taken into account.
This was done using the integrated interaction phase $\Phi_0=\int\Omega(t)^4/(8|\Delta|^3)\,\mathrm{d}t$ in the theoretical models, where $\Omega(t)$ denotes the instantaneous Rabi coupling which saturates after a finite rise time at a value $\Omega_s$. The temporal shape of the pulses is captured by two time constants and determined by a combination of the rise time of the acousto optic modulator and the bandwidth of the intensity stabilisation used to generate the pulses.\\
For the Ramsey fringe measurement shown in Fig. 1d in the main text, we used a sequence without the intermediate $\pi$ echo pulse and a single dressing phase of variable time $t$ (Fig.~\ref{fig:S1}a).

\section{Collective longitudinal field and Rabi frequency calibration}\label{SecCollField}

\begin{figure*}
  \centering
  \includegraphics[width=0.9\textwidth]{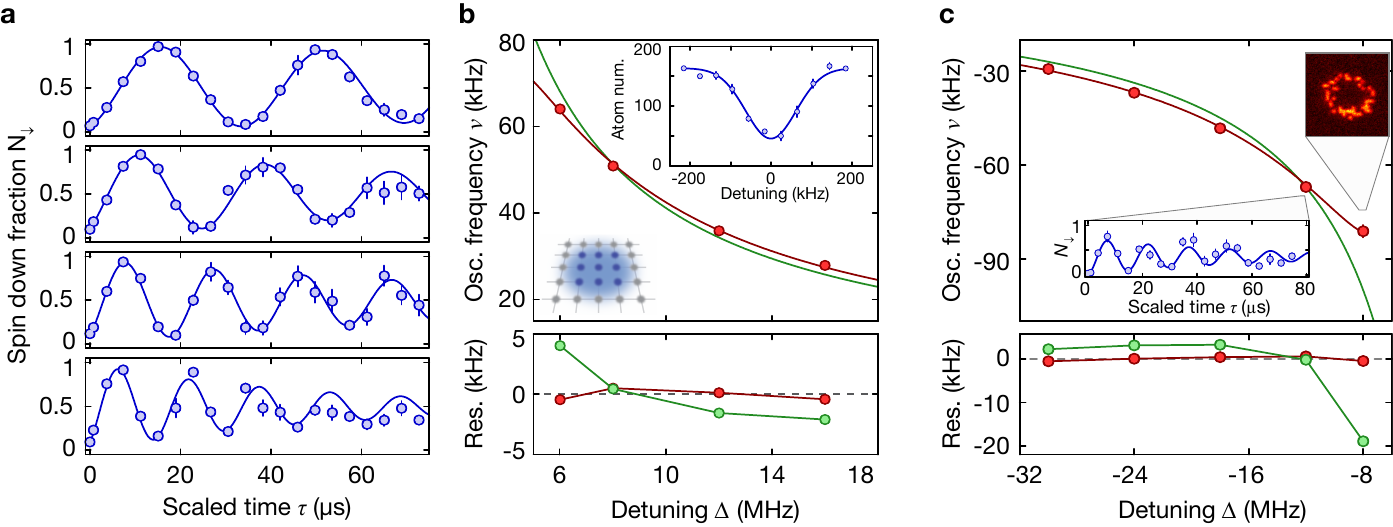}%
  \caption{\textbf{Extraction of the collective longitudinal field for $J=1/2$ and $J=3/2$.}
  	  \textbf{a}, Fraction of detected spin down atoms in a central elliptical region of interest with a radius of $0.7$ times the average Mott insulator radius for varying detuning ($\Delta/2\pi=16, 12, 8, 6\,\mathrm{MHz}$ from top to bottom) after Ramsey sequence without echo. 
  	  The solid blue line shows a fit of a damped sinusoidal oscillation to extract the frequency $\nu(\Delta)$. 
  	  The errorbars denote the standard error of the mean
      \textbf{b}, The upper panel shows the extracted dependence of the $\nu$ on detuning $\Delta$ (red points), including a fit with and without $\DelCol$ (red and green solid lines).
    The errorbars mark the $1\sigma$ confidence interval of the fit.
  The inset shows a high resolution Rydberg resonance measurement by optical pushout of ground state atoms (blue points) with a Gaussian fit (blue solid line).
  The pictogram in the lower left corner illustrates the collective longitudinal field shift $\DelCol$.
  Errorbars on the datapoints denote the standard error of the mean.
  The lower panel shows the residuals of the two different fits in the upper panel.
  The zero position is marked by the grey dashed line.
  The lines are guides to the eye.
  	\textbf{c}, Same analysis as in b for Ramsey oscillations in a spin system dressed to the Rydberg state $31P_{3/2}$ $m_J=-3/2$.
  	Contrary to b, here the oscillation frequency is negative due to the red detuning $\Delta<0$.
  The insets show an exemplary time trace of the mean fraction of spin down atoms $N_{\downarrow}$ for $\Delta/2\pi=-12\,$MHz and a single realisation of a spin down atom density distribution for one specific time of the oscillation with $\Delta/2\pi=-8\,\mathrm{MHz}$.
  The ring shaped density distribution illustrates the influence of the system boundary on the collective field.
  The mean atom numbers shown in this graph are obtained by averaging the result of $5-10$ experimental shots.} 
\label{fig:S2} 		
\end{figure*}

Next to the direct long range spin-spin interaction, Rydberg-dressing induces a collective longitudinal field for each spin due to the interaction with all its neighbours within the interaction range, c.f. Eq.~1 in the main text.
This shift can be directly extracted from the data by performing a Ramsey experiment, which is sensitive to single particle shifts (Fig.~\ref{fig:S2}).
The frequency of the Ramsey fringes shown here is dominated by the dressing laser induced single particle light shift 
\begin{equation}\label{eq:shift}
\delta(\Delta)=-\frac{\Delta}{2}+\frac{1}{2}\sqrt{\Omega^2+\Delta^2}\approx\frac{\Omega^2}{4\Delta}.
\end{equation}
We repeated such measurements for different detunings $\Delta$ and observed the expected increase in oscillation frequency $\nu$, extracted via an exponentially damped sinusoidal fit from the data, with decreasing $\Delta$ (Fig.~\ref{fig:S2}b).
However, a fit of the data $\nu(\Delta)$ with the single particle formula $\delta(\Delta)$ and the Rabi frequency $\Omega$ as a free parameter shows a clear systematic residual, which is removed when taking into account the mean field shift $\DelCol$ (cf. Eq.~1 in main text).
The latter is achieved with the relation $\nu(\Delta)=\delta(\Delta)-N_{\text{eff}}\frac{\Omega^4}{16\Delta^3}$ and the effective atom number $N_{\text{eff}}$ contributing to the shift.


Leaving $\Omega$ and $N_{\text{eff}}$ as fit parameters, we obtained the best agreement for $N_{\text{eff}}=11(2)$ particles within the interaction range, which agrees well with the simple estimate $\pi R_c^2=12.5$ for the $31P_{1/2}$ state with a cut-off radius of $R_c/\alat=2.02$.
At the same time this fit provides a calibration value for the Rabi coupling, yielding $\Omega_s/2\pi=1.33(7)\,\mathrm{MHz}$.
The error of $5\%$ both takes into account pulse to pulse and day to day fluctuations.
In the preceding analysis, we take the time dependence of $\Omega$ discussed in the previous section into account by scaling the time axis as $\tau=\int\Omega^2(t)\,\mathrm{d}t/\Omega_s^2$, where we normalise to the asymptotic value $\Omega_s$ reached when the dressing light power has reached its maximum.
This rescaling is suggested by the $\Phi\propto\Omega^2$ dependence of the phase $\Phi$ due to the dominating light shift. 
The rescaling factor of the time axis amounts to approximately $0.85$ for short dressing times but for longer times between $20$ and $80\,\mu$s it merely leads to a compression of the time axis by a factor of $0.95$.
Slight uncertainties in the saturation value $\Omega_s$ and hence the compression factor do not alter the results of the quantitative analysis of Fig.~\ref{fig:S2} within the errorbars.
The analysis of the collective longitudinal field relies also on the accurate knowledge of $\Delta$ and hence the resonance position.
Precise spectroscopy enabled the measurement of high resolution resonance curves, shown in the inset of Fig.~\ref{fig:S2}, from which the line center was extracted with an uncertainty of less than $10\,\mathrm{kHz}$. This in combination with slight drifts during the day determines the experimental uncertainty of $30\,\mathrm{kHz}$ for the detuning $\Delta$ assumed throughout the paper.
Furthermore, from the collective longitudinal field shift one can determine if the induced long range interaction is attractive, as in our case, or repulsive: The observed $\nu(\Delta)$ decreases due to the interactions if the sign of the interaction potential differs from the sign of the detuning.

\begin{figure*}
\centering
  \includegraphics[width=0.9\textwidth]{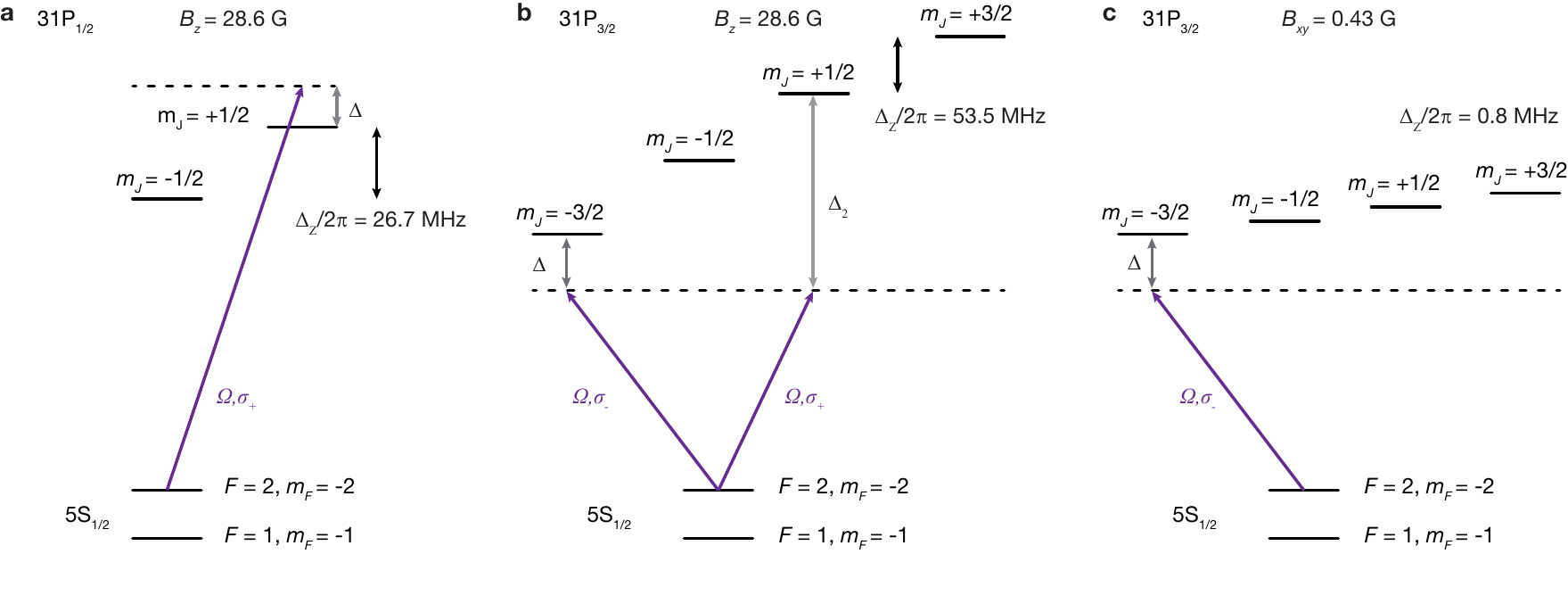}%
  \caption{ \label{fig:S3} \textbf{Level structure relevant for dressing to $31P_{1/2}$ and $31P_{3/2}$.}
  \textbf{a}, Dressing configuration for coupling to $31P_{1/2}$ with the magnetic offset field of strength $B_{z}=28.6\,$G in $z$-direction, yielding a differential Zeeman splitting of $\Delta_Z/2\pi=26.7\,\mathrm{MHz}$ between the $m_J$ states.
  The detuning was chosen to be $\Delta/2\pi=6\,\mathrm{MHz}$ to the blue of $m_J=+1/2$, which is coupled by the $\sigma_+$ polarisation component of the dressing laser.
  \textbf{b}, Dressing configuration for coupling to $31P_{3/2}$ with the same magnetic field as in a, yielding a differential Zeeman splitting of $\Delta_Z/2\pi=53.5\,\mathrm{MHz}$ between the $m_J$ states.
  The detuning is chosen to the red of $m_J=-3/2$ to be $\Delta/2\pi=-6\,\mathrm{MHz}$, which is coupled by the $\sigma_-$ polarisation component of the dressing laser.
  Coupling to $m_J=+1/2$ with the equally strong $\sigma_+$ polarisation component can be neglected for the dressing potential due to the additional Zeeman detuning $\Delta_2$ and the strong scaling of the dressing potential $U(0)\propto1/\Delta^3$.
  \textbf{c}, For demonstrating anisotropic interactions, the magnetic field was set in the $x-y$ plane of the atoms, aligned with the direction of the excitation laser.
  In the field of strength $B_{xy}=0.43\,\mathrm{G}$, the Zeeman states are split by $\Delta_Z/2\pi=802\,\mathrm{kHz}$.
  Using $\sigma_-$ polarisation, only the $m_J=-3/2$ state is optically coupled.
  The detuning for the anisotropy measurement was chosen to be $\Delta/2\pi=-12\,\mathrm{MHz}$ with respect to the $31P_{3/2}$, $m_J=-3/2$ state.	 
  }
\end{figure*}
We repeated the same measurement also for the state $31P_{3/2}$, which is dressed to the ground state via the $\ket{F=2, m_F=-2}$ to $\ket{J=3/2, m_J=-3/2}$ transition.
Fig.~\ref{fig:S2} summarises the result for a region of interest spanning an elliptical central area of the Mott insulator containing approximately half the number of atoms of the full sample.
As expected, the absolute value of the Ramsey oscillation frequency increases with decreasing $|\Delta|$.
The sign of the oscillation frequency indicates a decreasing transition frequency due to the dressing light.
However, similarly to the observation for $31P_{1/2}$, the single parameter fit with $\Omega$ as the only free parameter does not capture the dependence of the oscillation frequency on $\Delta$. 
The fit residual vanishes only when taking into account the collective longitudinal field shift $\DelCol$.
From the fit, we conclude that $N_{\text{eff}}=19(1)$ particles contribute to this shift.
The deviation from the expected value of $R_c^2\pi\approx3^2\pi\approx 28.3$ could be due to a reduced mean density due to loss processes or residual finite size effects, as particles on the edge of the initially prepared sample experience a smaller collective shift, leading to a decreased detected average shift.
From the same fit, we extract a Rabi coupling of $\Omega_s/2\pi=1.9(1)\,\mathrm{MHz}$ to the Rydberg state.
For the measurements shown in Fig. 3a in the main text, the optical power is reduced and we work with a Rabi frequency of $\Omega_s/2\pi=1.16(6)\,\mathrm{MHz}$.
The same procedure was applied to calibrate the Rabi frequency for the anisotropy measurement shown in Fig. 3b in the main text and we obtained a value of $\Omega_s/2\pi=2.45(12)\,\mathrm{MHz}$.

\section{Interaction potentials induced by Rydberg-dressing\label{SecInteractionPotentials}}
\subsection{Single particle excitation scheme \label{SecSingleParticleExcitation}}

Fig.~\ref{fig:S3} shows the laser and state configurations relevant for the data shown in Figs.~2-4 in the main text. In all cases, we applied a magnetic field to lift the Zeeman degeneracy of the $\ket{F=2}$ ground state manifold. We ensured that the corresponding Zeeman splitting greatly exceeds the Rabi frequency of the applied microwave pulses in order to isolate the $\ket{F=1,m_F=-1}\rightarrow\ket{F=2,m_F=-2}$ transition as the spin system. Therefore the orientation of the magnetic field defines the quantisation axis for the effective spins.

The $31P_{1/2}$ Rydberg state (Fig.~\ref{fig:S3}a) is coupled by the $\sigma_{+}$ polarisation component of the excitation beam, which propagated in the $x-y$ plane, at an angle of $90^\circ$ with respect to the magnetic field $B_{z}=28.6\,\mathrm{G}$ applied along the $z$-direction (c.f. Fig.~1b). Hence, the effective spins were defined orthogonal to the plane of the lattice. The two Rydberg Zeeman sublevels $m_J=-1/2$ and $m_J=+1/2$ are split in this field by $\Delta_Z/2\pi=26.7\,\mathrm{MHz}$. Due to optical selection rules, only the state $\ket{J=1/2, m_J=+1/2}$ is coupled to the ground state. We chose a blue detuning of $\Delta/2\pi=6\,\mathrm{MHz}$ to operate in the weak dressing regime.

For the data shown in Fig.~3a of the main text, the ground state $\ket{F=2, m_F=-2}$ is coupled to the $31P$ $\ket{J=3/2, m_J=-3/2}$ Rydberg state with a red detuning of $\Delta/2\pi=-6\,\mathrm{MHz}$ (Fig.~\ref{fig:S3}b). Since we used a linearly polarised excitation beam with the polarisation axis lying in the plane of the atomic lattice, the $\ket{J=3/2, m_J=+1/2}$ state is optically coupled as well. However, the large Zeeman spitting $\Delta_Z/2\pi=53.5\,\mathrm{MHz}$ of the Rydberg manifold due to the applied magnetic field, $B_{z}=28.6\,\mathrm{G}$, rendered this coupling negligible. Again the effective spins formed by the two ground states were defined perpendicular to the optical lattice along the magnetic field axis.

For the data shown in Fig.~3b of the main text, we used a smaller magnetic field $B_{xy}=0.43\,\mathrm{G}$ that was aligned in the plane of the atoms along the diagonal of the square lattice set by the optical trapping fields. Optical coupling was provided by the $\sigma_-$-polarised Rydberg-excitation laser propagating along the magnetic field direction. Therefore, in this case, the quantisation axis of the effective spins was defined in the plane of the optical lattice.
\subsection{Calculation of the Rydberg-Rydberg atom interaction potential}\label{SecBarePotentials}

For the calculations of the Rydberg-Rydberg interaction potentials we consider two atoms with an interatomic separation vector $\bf R$. The uncoupled ground states do not participate in the Rydberg-dressing such that it suffices to consider the ground state $\ket{\rm g} = \ket{F = 2, m_F = -2}$ and the manifold of Rydberg states $\ket{\Rydblvl} = |n L J m_J\rangle$, with principal quantum number $n$, orbital angular momentum $L$, and total angular momentum $J$ with its projection $m_J$ along the quantisation axis. Introducing pair states, $\ket{\alpha\beta}$ ($\alpha,\beta={\rm g},\Rydblvl$), the corresponding atomic Hamiltonian can be written as 
\begin{equation}\label{eq:HA}
\hat{H}_{\rm A}=-\sum_\Rydblvl \Delta_\Rydblvl(\ket{{\rm g}\Rydblvl}\bra{{\rm g}\Rydblvl}+\ket{\Rydblvl{\rm g}}\bra{\Rydblvl{\rm g}})-\sum_{\Rydblvl,\Rydblvl^\prime} (\Delta_\Rydblvl+\Delta_{\Rydblvl^\prime})\ket{\Rydblvl\Rydblvl^\prime}\bra{\Rydblvl\Rydblvl^\prime}  
\end{equation}
where $\Delta_\Rydblvl$ denotes the laser detuning with respect to a given Rydberg state, with $\Delta_{\Rydblvl_0}=\Delta$ being the detuning of the targeted Rydberg state, i.e., $\ket{\Rydblvl_0}=\ket{31 P J m_J}$ for the measurements presented in this work. Note that the detunings $\Delta_\Rydblvl$ also depend on $m_J$ through the Zeeman shift by the external magnetic field, whose direction we chose to define the quantisation axis.

For these Rydberg states the minimum atomic distance set by the lattice constant is sufficiently large to justify a leading-order description of the electrostatic atomic interaction in terms of the dipole-dipole interactions, in atomic units, 
\begin{equation}\label{eq:Vdd}
\hat{V}_{\rm dd}=\sum_{\Rydblvl,\bar{\Rydblvl},\Rydblvl^\prime,\bar{\Rydblvl}^\prime}\frac{{\bf{d}}_{\Rydblvl,\bar{\Rydblvl}} \cdot {\bf{d}}_{\Rydblvl^{\prime},\bar{\Rydblvl}^{\prime}}}{R^3} - \frac{3({\bf{d}}_{\Rydblvl,\bar{\Rydblvl}}\cdot {\bf R})({\bf{d}}_{\Rydblvl^\prime,\bar{\Rydblvl}^\prime}\cdot {\bf R})}{R^5}\ket{\bar{\Rydblvl}\bar{\Rydblvl}^\prime}\bra{\Rydblvl\Rydblvl^\prime},
\end{equation}
that couple atomic pair states with corresponding one-body transition matrix elements ${\bf d}_{\Rydblvl,\bar{\Rydblvl}}=-\bra{\Rydblvl}{\bf r}\ket{\bar{\Rydblvl}}$.

\begin{figure*}
\centering
  \includegraphics[width=0.9\textwidth]{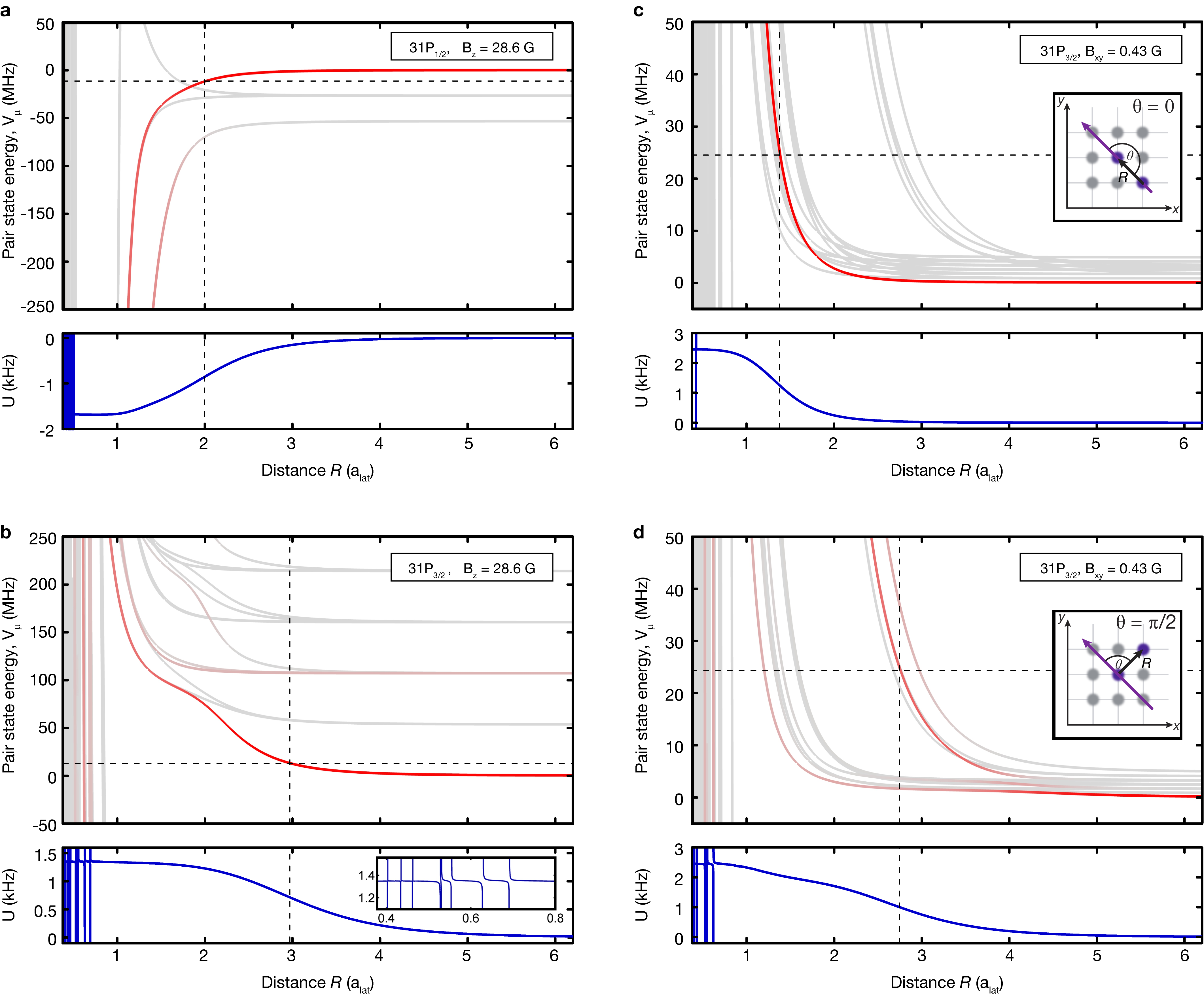}
  \caption{\textbf{Calculated interaction potentials.}
  {\bf a} and {\bf b}, Upper panels: Rydberg-Rydberg atom potential curves relevant for coupling to $31P_{1/2}$ and $31P_{3/2}$ in a magnetic field of $B_z=28.6\,\mathrm{G}$ aligned with the $z$ axis. Each of the shown asymptotes belongs to a combination of $m_J$ such that the asymptotic splitting corresponds to twice the Zeeman splitting $\Delta_Z$.
  The intensity of the red colouring indicates the relative coupling strength $\bar{\alpha}_{e_0{\rm g}}^{(\mu)}$ [Eq.(\ref{eq:amu})] for a given potential curve $V_\mu$. Lower panels: Resulting interaction potential $U$ between two Rydberg-dressed ground state ($\ket{\tilde{g}}$) atoms. The dashed vertical line marks the the range $R_c$ of the effective interaction, determined by the detuning and interaction potential of the most strongly coupled pair-state according to $|V_\mu(R_{\rm c})|=2|\Delta|$.
Inset of b provides a closer view at small $R$.
{\bf c} and {\bf d}, Same plots as a and b, for a magnetic field $B_{xy} = 0.43\,\mathrm{G}$ aligned in-plane with the atoms, and the atomic separation vector ${\bf R}$ either aligned with the magnetic field c, or perpendicular to it d, as depicted in the insets. The dressing-induced interactions have been obtained for Rabi frequencies of a $\Omega_s/2\pi=1.33\,$MHz, b $\Omega_s/2\pi=1.25\,$MHz and c, d $\Omega_s/2\pi=2.45\,$MHz and detunings of a $\Delta/2\pi=6\,$MHz, b $\Delta/2\pi=-6\,$MHz and c, d $\Delta/2\pi=-12\,$MHz.
  }
   \label{fig:S4} 
\end{figure*}

We have diagonalised $\hat{H}_A+\hat{V}_{\rm dd}$ using a large basis set of atomic pair states around the laser coupled asymptote, $\ket{\Rydblvl_0\Rydblvl_0}$, ensuring convergence for distances relevant to our experiments. This yields a set of potential curves $V_{\mu}({\bf R})$ (Fig.~\ref{fig:S4}) corresponding to a given molecular eigenstate 
\begin{equation}\label{eq:MolecularEigenstates}
\ket{\mu({\bf R})} = \sum_{\Rydblvl\Rydblvl^\prime} c^{(\mu)}_{\Rydblvl\Rydblvl^\prime}({\bf R}) \ket{\Rydblvl\Rydblvl^\prime}.
\end{equation}
which depend on distance and orientation ($\bf R$) of the interacting atomic pair through the coefficients $c^\mu_{\Rydblvl\Rydblvl'}({\bf R})$. The effective potential generated by Rydberg-dressing is determined by both the potential curves, $V_\mu$, as well as the associated molecular states, $\ket{\mu}$.\\

\subsection{Calculation of the ground state interaction potential induced by Rydberg-dressing}\label{SecDressedPotentials}

We now turn our attention to the laser coupling
\begin{widetext}
\begin{equation}
\hat{H}_{\rm L}=\frac{\Omega}{2}\sum_\Rydblvl\alpha_\Rydblvl \ket{{\rm gg}}\left(\bra{{\rm g}\Rydblvl}+\bra{\Rydblvl{\rm g}}\right)+\frac{\Omega}{2}\sum_{\Rydblvl,\mu}\bar{\alpha}_{\Rydblvl{\rm g}}^{(\mu)}\left(\ket{{\rm g}\Rydblvl}+\ket{\Rydblvl{\rm g}}\right)\bra{{\mu}({\bf R})}+{\rm h.c.}
\end{equation}
\end{widetext}
where the coefficients $\alpha_{\Rydblvl}$ account for the different laser-coupling strength of a given Rydberg state $\ket{\Rydblvl}$ relative to the Rabi frequency $\Omega$ of the target state $\ket{\Rydblvl_0}$, for which $\alpha_{\Rydblvl_0}=1$. The first term describes the laser-coupling of the two-atom ground state to the singly excited pair state, while the second term corresponds to the subsequent coupling to a doubly excited molecular Rydberg state $\ket{\mu}$. The corresponding coefficient 
\begin{equation}\label{eq:amu}
\bar{\alpha}_{\Rydblvl{\rm g}}^{(\mu)}=\sum_{\Rydblvl^\prime}\alpha_{\Rydblvl^\prime}c^{(\mu)}_{\Rydblvl\Rydblvl^\prime}
\end{equation}
accounts for the laser-coupling to a given molecular state relative to $\Omega$ and follows from the calculated form of $\ket{\mu({\bf R})}$ given in Eq.(\ref{eq:MolecularEigenstates}). The red colouring of the molecular potential curves shown in Fig.~\ref{fig:S4} corresponds to $\bar{\alpha}_{\Rydblvl_0{\rm g}}^{(\mu)}$ and is thus indicative of the laser-coupling to a given curve via the target state $\ket{\Rydblvl_0}$.

The distance dependent Rabi frequencies $\bar{\alpha}_{\Rydblvl{\rm g}}^{(\mu)}\Omega$ in conjunction with the potential curves $V_\mu$ give rise to a distance dependent light shift $\delta E({\bf R})$ of the dressed ground state $\ket{\tilde{\rm g}\tilde{\rm g}}$. The resulting effective potential due to Rydberg-dressing thus emerges as the difference of the two-atom light shift and the asymptotic values, $U({\bf R})=\delta E({\bf R})-\delta E(|{\bf R}|\rightarrow\infty)$.
We obtain this potential by numerically diagonalising the full Hamiltonian $\hat{H}_{\rm A}+\hat{V}_{\rm dd}+\hat{H}_{\rm L}$.
Fig.~\ref{fig:S4} shows the potentials obtained for dressing to the $31P_{1/2}$ (Fig.~\ref{fig:S4}a) and $31P_{3/2}$ (Figs.~\ref{fig:S4}b-d) states, respectively. Qualitatively, the potentials exhibit the expected soft-core shape, with a $1/ R^6$ tail at large $R$ saturating to a finite plateau value near $R = 0$, while quantitative differences arise from the more complex potential structure compared to the assumption of simple van der Waals interactions between Rydberg atoms. Our exact results reveal sharp resonances in the plateau region which arise from resonance with molecular potentials $V_\mu$ that feature a finite value of $\bar{\alpha}_{\Rydblvl{\rm g}}^{(\mu)}$. Since $\bar{\alpha}_{\Rydblvl{\rm g}}^{(\mu)}\ll1$ these resonance are generally very narrow, but can nevertheless cause resonant Rydberg-state excitation which leads to increased losses. Through a careful choice of the targeted Rydberg states such potential losses are avoided by our optical lattice.

\begin{figure*}
  \centering
  \includegraphics[width=0.9\textwidth]{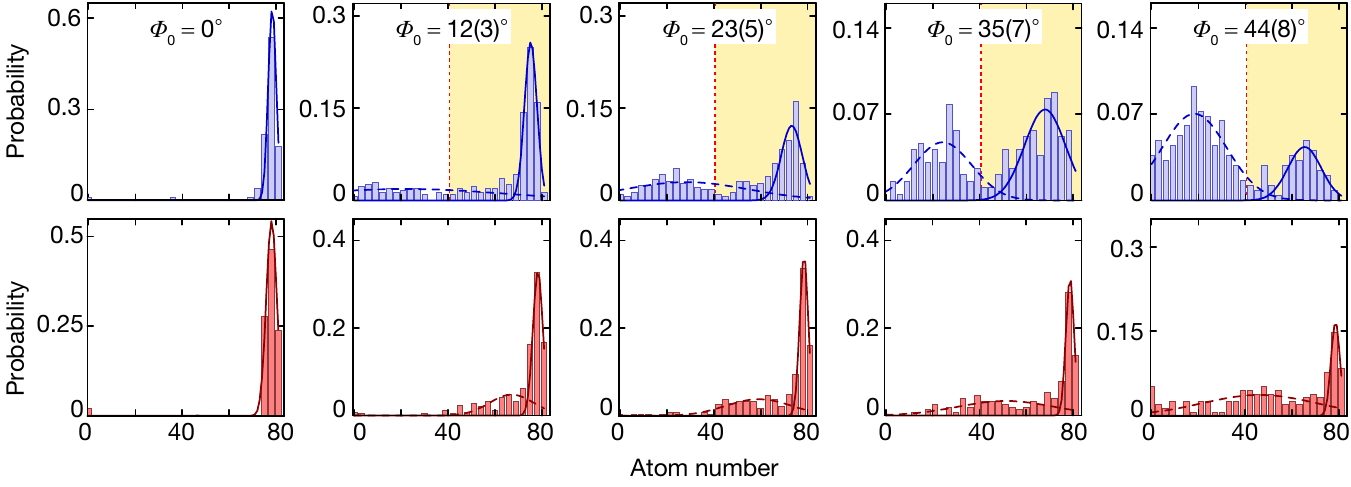}%
 \caption{ \label{fig:S5} \textbf{Dynamics of atom number histograms.}
  We show atom number histograms for the varying interaction phases $\Phi_0$ (as in Fig. 2 in the main text) for detection with and without spin sensitive detection (upper and lower row, respectively).
  The dashed (solid) lines in the histograms are results of fits with a sum of Gaussian peaks for the left (right) peaks.
  The red dashed line marks the post-selection threshold, the yellow shaded area indicates the data used to calculate $\gtwo$ shown in Fig. \ref{fig:S6}.
  The histograms were extracted from the atom number in a region of interest of $9\times9$ central sites to suppress the effect of edge fluctuations of the sample. } 
\end{figure*}

Figs.~\ref{fig:S4}a and b show the potentials used for the calculations of Fig. 2 and 3a of the main text. As can be seen, changing the angular momentum from $J=1/2$ to $J=3/2$ allows to switch the sign of the most relevant Rydberg-Rydberg atom potential curve and thereby permits to generate attractive (Fig.~\ref{fig:S4}a) and repulsive (Fig.~\ref{fig:S4}b) effective spin interactions. Moreover, the absolute value of the Rydberg-Rydberg atom interaction is significantly enhanced for $J=3/2$ leading to a longer range of the corresponding potential. In both of these cases we used a linearly polarised excitation laser whose propagation direction and polarisation vector lies in the plane of the atomic lattice. As this laser-field orientation breaks the rotational symmetry, one might expect an anisotropy of the dressing-induced interactions. However, we have applied a large magnetic field, $B_z=28.6\,\mathrm{G}$, orthogonal to the lattice plane which causes a Zeeman shift that greatly exceeds both the Rabi frequency, $\Omega$, and the detuning $\Delta$ of the Rydberg-excitation laser (Fig.~\ref{fig:S3}). Hence, the magnetic field dictates the underlying symmetry and, thereby, forces the induced interactions to be isotropic which is directly reflected in our correlation measurements shown in Figs.~2a and 3a of the main text.

This situation changes dramatically for the measurements shown in Fig. 3b of the main text, where we used a smaller magnetic field, $B_{xy}=0.43\,\mathrm{G}$, now aligned in the plane of the atoms along the propagation axis of the circularly polarised excitation beam. As shown in Figs.~\ref{fig:S4}c and d, this leads to strongly anisotropic interactions. Note that the actual potential curves $V_{\mu}$ remain nearly isotropic, just acquiring a weak anisotropy due to the applied magnetic field. However, the composition of the molecular states in Eq. (\ref{eq:MolecularEigenstates}) is very sensitive to the orientation of the distance vector ${\bf R}$ with respect to the quantisation axis, which is fixed relative to the wave vector of the excitation laser and the magnetic field direction. As a result the laser effectively couples to different curves depending on the orientation, as indicated by the red colouring in Figs.~\ref{fig:S4}c and d. Eventually, this leads to the strong anisotropy of the calculated dressing-induced interactions and our measured correlation functions shown in Fig. 3b of the main text.

These examples demonstrate the high degree of tunability via the choice of targeted Rydberg state as well as by controlling the strength and orientation of the applied magnetic field and excitation laser.

\section{Correlated loss and post-selection}\label{SecPostSelection}


\begin{figure*}
  \centering
  \includegraphics[width=0.9\textwidth]{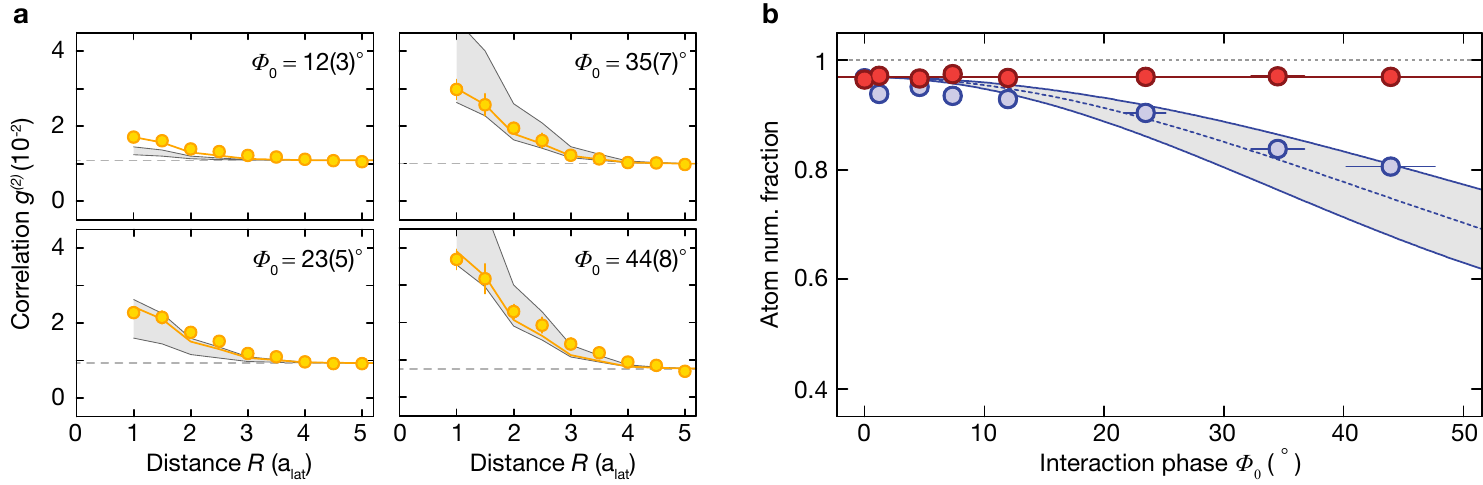}%
  \caption{ \label{fig:S6} \textbf{Dynamics and correlations on post-selected data.}
  \textbf{a}, Azimuthally averaged $\gtwo$-correlations with post-selection for increasing interaction phases $\Phi_0$ (yellow datapoints).
 The grey shaded area denotes the region of uncertainty for the measured spin correlation due to the $5\%$ fluctuations and uncertainty of the Rabi frequency $\Omega_s$.
  The solid yellow line represents a best fit of the correlation amplitude adjusting $\Omega$, shifted by the constant offset $\gtwo_{\infty}$ (grey dashed line, see main text for details)
    \textbf{b}, Fraction of atoms extracted from the Gaussian fit to the measured atom number histograms (Fig. \ref{fig:S5}) in a region of interest of $9\times9$ sites with and without spin selective detection (blue and red points).
    The grey shaded area indicates the expected fraction of spin down atoms assuming fully coherent Ising dynamics for a Rabi frequency range $\Omega_s/2\pi=1.33(7)\,\mathrm{MHz}$ with the center indicated by the dashed blue line.
    The red solid line indicates the initial state filling of $97\%$ in our experiment.
    The grey dashed line marks the expected fraction of $1$ for a perfectly filled initial state.}
\end{figure*}

To obtain deeper insight into the loss processes observed in the experiment, we analyse the number histograms underlying the mean atom numbers shown in Fig. 4 in detail on a region of interest of $9\times9$ sites in the center of the initial Mott insulator.
The set of atom number histograms is shown in Fig.~\ref{fig:S5}.
The observed long tail for short times or the bimodality for later times are not expected from the coherent theory without dissipation and support the hypothesis of a loss process which is correlated in the sense that a single trigger event leads to a significant loss of atoms in the whole ensemble.
Assuming this kind of process one can explain the bimodal histograms as well as the spatial offset of the correlation measurements.
In order to analyse the measurement results independently of this loss process, we fit the histograms with a sum of two Gaussians (Fig.~\ref{fig:S5}).
The strong bimodality for long times allows to interpret the right Gaussian peak as mostly consisting of measurement outcomes where no trigger event has happened.
This is confirmed by using the position of this peak as an estimate for the mean atom number and the good agreement with the coherent theory without dissipation, both with and without spin selective detection (Fig.~\ref{fig:S6}), and a similar measurement could reveal the scaling of interaction induced dephasing \cite{Mukherjee2015}.
Notably, the data without spin selective detection shows no significant decay up to the largest times, which is in accordance with the lifetime of $2.2\,\mathrm{ms}$ expected if the ideal dressing approximation is applied.
Focusing on the left peaks, the ratio of the center positions for the measurements with and without resolving the spin equals approximately $1/2$ for interaction phases $\Phi_0\geq23(5)\,\mathrm{^{\circ}}$, which indicates that before the final $\pi/2$-pulse one spin component is fully eliminated from the sample by the loss process.
We can furthermore compare the integrated weight within the right fitted Gaussian peak for detection with or without spin-resolved measurement.
Here, we obtain very similar weights for the two cases, which allows to conclude that the acquired phase in events without trigger event is not significantly affected by the loss process.

Exploiting the strong bimodality of the atom number histograms also allows to analyse the spin-spin correlation function $\gtwo$ (Fig. 2) on post-selected datasets.
If only events are kept where the detected fraction of atoms within the region of interest of $9\times9$ sites is larger than $1/2$, we obtain the spin-spin correlation shown in Fig.~\ref{fig:S6}.
Especially for the small interaction phase $\Phi_0=12(3)\,\mathrm{^\circ}$, the bimodality of the histogram has not yet fully developed, which hints at a finite time scale associated with the loss of the sample after the initial black-body trigger event. While this introduces some ambiguity into the post-selection procedure, setting a cutoff fraction of $1/2$ appears reasonable since a $\ket{\downarrow}$-fraction lower than this value can not be expected from the purely unitary dynamics at early times.
For the largest two interaction phases $\Phi_0\geq35(7)\,\mathrm{^\circ}$, however, the post-selection mainly retains events in the high atom number peak, which predominantly result from the unitary dynamics, undisturbed by dissipative processes.
Comparing the $\gtwo$ amplitude for increasing interaction phase $\Phi_0$ with the expected result for the ideal analytic result without loss and taking into account the $5\%$ variation of the Rabi frequency of $\Omega_s/2\pi=1.33(7)\,\mathrm{MHz}$ we obtain quantitative agreement.
Whereas the correlation amplitude is not changed significantly apart from an increase for the largest interaction phases $\Phi_0$, the constant global offset $\gtwo_{\infty}$ is strongly reduced, as expected when the loss events with a correlation range exceeding the size of the whole sample are removed.

\begin{figure}
  \centering
  \includegraphics[width=0.45\textwidth]{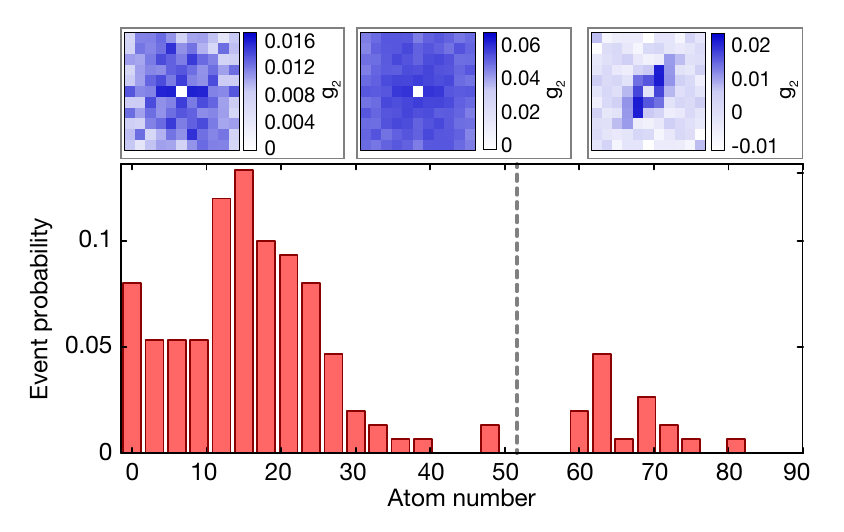}%
  \caption{\textbf{Atom number histogram for anisotropy measurement.}
  The number histogram of atoms remaining after the Ramsey sequence including spin echo (Fig. 3b) shows strong bimodality.
  Evaluation of the spin-spin correlation on the events in the left peak of the histogram, as well as on the full dataset show no significant structure but a large offset $\gtwo_{\infty}$ (left and middle panel above the figure), whereas post-selection on the events in the right peak, bears out the anisotropic correlation (right panel above figure).
}
 \label{fig:S7} 
\end{figure}

For the $31P_{3/2}$ anisotropy data shown in Fig. 3b in the main text the anisotropic correlation signal nearly disappears in the background when we analyse the data without post-selection (Fig.~\ref{fig:S7}).
As in the $31P_{1/2}$ case, this background stems from the atom loss initiated by black-body radiation. However, the post-selection filters the corresponding events effectively and yields a vanishing offset for $\gtwo$ in quantitative agreement with the fully coherent dynamics predicted by our theory (Fig.~3b). Note that Fig.~3b shows the only post-selected dataset in the main text.

\section{Theoretical description of the spin dynamics}\label{SecTheoSpinDynamics}

Although the applied pulse sequences (Fig.~\ref{fig:S1}) can generate sizeable spin correlations and entanglement, one can nevertheless formulate an exact solution for the final many-body state. This is possible because the dynamics induced by the microwave pulses and Rydberg-dressing stages separately permit an analytical description. 

The time evolution operator $\hat{U}_{\pi/2}=\bigotimes_i \hat{U}_{\frac{\pi}{2}}^{(i)}$ of the microwave $\pi/2$-pulse factorizes into one-body operators acting on particle $i$
\begin{equation}
\hat{U}_{\frac{\pi}{2}}^{(i)} = \frac{1}{\sqrt{2}}\matfour{1}{-\mathrm{i}}{-\mathrm{i}}{1}\;,
\end{equation}
transforming each spin state as $\ket{\downarrow} \to (\ket{\downarrow} - \mathrm{i}\ket{\uparrow})/\sqrt{2}$ and $\ket{\uparrow} \to (\ket{\uparrow} - \mathrm{i}\ket{\downarrow})/\sqrt{2}$. Equivalently, we have $\hat{U}_{\pi}=\bigotimes_i \hat{U}_{\pi}^{(i)}$
\begin{equation}
\hat{U}_{\pi}^{(i)} = \hat{U}_{\frac{\pi}{2}}^{(i)}\hat{U}_{\frac{\pi}{2}}^{(i)}=\matfour{0}{-\mathrm{i}}{-\mathrm{i}}{0}\;.
\end{equation}
The Hamiltonian that governs the spin evolution during the Rydberg-dressing stage is given by Eq.(1) of the main text. For the analysis below it is more convenient to reformulate the Hamiltonian 
\begin{equation}\label{EqHamiltonianInteractionPhase}
\hat{H}_{\rm dr} = \sum_{i<j} U_{ij}(t) \hat{\sigma}_{\uparrow\uparrow}^{(i)}\hat{\sigma}_{\uparrow\uparrow}^{(j)} + \sum_i \delta(t) \hat{\sigma}_{\uparrow\uparrow}^{(i)}
\end{equation}
in terms of projection operators $\hat{\sigma}_{\uparrow\uparrow}^{(i)}=\ket{\uparrow}_i\bra{\uparrow}$ for the $i^{\rm th}$ atom. The single-atom energy shift is given by Eq.(\ref{eq:shift}) and $U_{ij}$ denotes the time dependent Ising spin interaction generated by the Rydberg-dressing pulse. The corresponding time evolution operator over the dressing period
\begin{equation}
\hat{U}_{\rm dr}=\bigotimes_i\exp\left(-\mathrm{i}\varphi \hat{\sigma}_{\uparrow\uparrow}^{(i)}-\mathrm{i}\sum_{j>i}\Phi_{ij} \hat{\sigma}_{\uparrow\uparrow}^{(i)}\hat{\sigma}_{\uparrow\uparrow}^{(j)}\right)
\end{equation}
induces correlated phase rotations while leaving the spin-state populations unchanged. Here we have defined the phases
\begin{equation}
\varphi =\int \delta(t) {\rm d}t\:,\quad\Phi_{ij} =\int U_{ij}(t) {\rm d}t
\end{equation} 
where the time integration is understood to extend over a single dressing pulse. Note that $\ket{\uparrow}$ refers to the dressed hyperfine ground state, assuming adiabatic following up a varying Rabi frequency $\Omega(t)$. In particular $\ket{\uparrow}=\ket{F=2,m_F=-2}\equiv\ket{g_\uparrow}$ if $\Omega=0$ during the microwave pulses.

\subsection{Ramsey spectroscopy}\label{SecTheoRamsey}

In our experiment we prepare all atoms in the $\ket{\downarrow}$-state, i.e. the initial many-body state is $\ket{\Psi(0)}=\ket{\downarrow\downarrow\ldots\downarrow}$. Application of the first $\pi/2$-pulse, thus creates a superposition state 
\begin{equation}
\hat{U}_{\frac{\pi}{2}}\ket{\Psi(0)}=\sum_{\sigma_1, \ldots, \sigma_N} f_{\sigma_1}f_{\sigma_2} \ldots f_{\sigma_N} \ket{\sigma_1 \sigma_2 \ldots \sigma_N},
\end{equation}
of all possible spin configurations, where $\sigma_i = \{\downarrow, \uparrow\}$ and $f_{\sigma_i}$ denotes the amplitude for atom $i$ of its state $\ket{\sigma_i}$, i.e. $f_{\downarrow} = 1/\sqrt{2}$ and $f_{\uparrow} = -\ii / \sqrt(2)$. The subsequent dressing stage does not affect the populations of these $N$-body state components but causes a state-dependent phase picked up by each component $\ket{\sigma_1 \sigma_2 \ldots \sigma_N}$. Applying the time evolution operator given above we can, therefore, write the many-body state at the end of the dressing stage as
\begin{equation}\label{eq:psi}
\ket{\Psi } = \sum_{\sigma_1, .., \sigma_N} \prod_{i} \br{f_{\sigma_i} \ee^{-\ii \varphi  s^{(\uparrow)}_{\sigma_i}}\ee^{-\ii \sum_j \Phi_{ij} s^{(\uparrow)}_{\sigma_i}s^{(\uparrow)}_{\sigma_j}}}\ket{\sigma_1 \sigma_2 \ldots \sigma_N},
\end{equation}
where $s^{(\uparrow)}_{\sigma_i}$ is defined as $\hat{\sigma}_{\uparrow\uparrow}^{(i)}\ket{\sigma_i}=s^{(\uparrow)}_{\sigma_i}\ket{\sigma_i}$, i.e., valued $s^{(\uparrow)}_{\sigma_i}=1$ if $\sigma_i = \ \uparrow$ and $0$ otherwise.

The Ramsey interferometer is closed by a final $\pi/2$ pulse, such that we can calculate any observable after the Ramsey sequence from
\begin{equation}\label{eq:exp}
\langle\hat{O}\rangle = \bra{\Psi }\hat{U}_{\frac{\pi}{2}}^{-1} \hat{O} \hat{U}_{\frac{\pi}{2}} \ket{\Psi }.
\end{equation}
For instance, for the spin projection operator $\hat{\sigma}_{\downarrow\downarrow}^{(i)}$, measured in our experiment, we have
\begin{equation}\label{eq:sdd}
\hat{U}_{\frac{\pi}{2}}^{-1} \hat{\sigma}_{\downarrow\downarrow}^{(i)} \hat{U}_{\frac{\pi}{2}} = \frac{1}{2} \left[\hat{\sigma}_{\downarrow\downarrow}^{(i)} + \hat{\sigma}_{\uparrow\uparrow}^{(i)} + \mathrm{i}(\hat{\sigma}_{\uparrow\downarrow}^{(i)} - \hat{\sigma}_{\downarrow\uparrow}^{(i)})\right].
\end{equation}
where $\hat{\sigma}_{\uparrow\downarrow}^{(i)}=\ket{\uparrow}_i\bra{\downarrow}$.
Combining Eqs.(\ref{eq:psi})-(\ref{eq:sdd}) we finally obtain for the Ramsey signal
\begin{equation}\label{EqRamseyCoherent}
\langle\hat{\sigma}_{\downarrow\downarrow}^{(i)}\rangle = \frac{1}{2} - \frac{1}{2} \mathrm{Re}\brs{\mathrm{e}^{-\mathrm{i} \varphi}\prod_{k \neq i} \br{\frac{1}{2} + \frac{1}{2}\ee^{-\ii \Phi_{ik}}}}.
\end{equation}

\begin{figure*}
\includegraphics[width=0.8\textwidth]{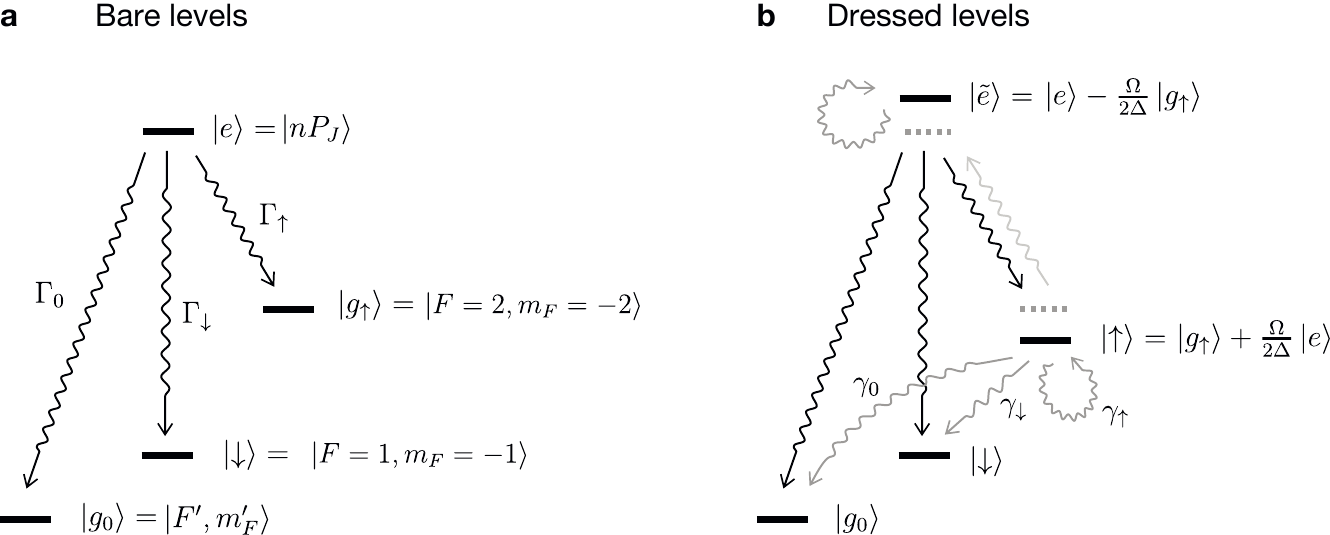}
\caption{\textbf{Spontaneous decay channels for bare and dressed atomic states.}
\textbf{a}, Schematic representation of the bare atomic levels, indicating decay rates from the Rydberg state. 
\textbf{b}, Dressed levels (up to first order perturbation theory): the $\ket{g_\uparrow}$ state is dressed with the Rydberg level $\ket{\Rydblvl}$, such the new eigenstates $\ket{\uparrow}$ and $\ket{\tilde{\Rydblvl}}$ will acquire a light shift. Additionally, new decay channels are opened up out of the dressed ground state $\ket{\uparrow}$ to all other states, including itself.} \label{fig:S8}
\end{figure*}

\subsection{Spin echo sequence}\label{SecTheoEcho}
Repeating the same steps for the spin echo sequence we can write the $N$-body wave function after the second dressing stage as
\begin{eqnarray}
\ket{\Psi} &=& \sum_{\sigma_1, .., \sigma_N } \prod_{i} \Big(\tilde{f}_{\sigma_i} \ee^{-\ii \left[\varphi^{(1)} s_{\sigma_i}^{(\downarrow)} + \varphi^{(2)}s_{\sigma_i}^{(\uparrow)}\right]} \nonumber\\
&&\times \ee^{-\ii \sum_j \left[ \Phi_{ij}^{(1)}s_{\sigma_i}^{(\downarrow)}s_{\sigma_j}^{(\downarrow)} + \Phi_{ij}^{(2)} s_{\sigma_i}^{(\uparrow)} s_{\sigma_j}^{(\uparrow)}\right]}\Big)\ket{\sigma_1 \sigma_2 \ldots \sigma_N},
\end{eqnarray}
where  $\tilde{f}_{\sigma_i}$ is the single particle amplitude after the $\pi$ pulse ($\tilde{f}_{\downarrow} = -1 / \sqrt{2}$ and $\tilde{f}_{\uparrow} = -\ii / \sqrt{2}$) and $s_{\sigma_j}^{(\downarrow)}=1-s_{\sigma_j}^{(\uparrow)}$. 
The phases $\varphi^{(1)}$ and $\varphi^{(2)}$ denote the total accumulated single particle phases of the first and second dressing pulse, respectively. Similarly, $\Phi_{ij}^{(1)}$ and $\Phi_{ij}^{(2)}$ denote the total accumulated interaction phases during the first and second dressing pulse, respectively. 
Using this expression in Eq.~(\ref{eq:sdd}) we obtain for the spin echo signal
\begin{equation}\label{EqEchoSignalCoherent}
\langle\hat{\sigma}_{\downarrow\downarrow}^{(i)}\rangle = \frac{1}{2} + \frac{1}{2}\mathrm{Re} \brs{\mathrm{e}^{\mathrm{i} [\varphi^{(1)} - \varphi^{(2)}]}\prod_{j\neq i} \frac{1}{2}\br{\mathrm{e}^{\mathrm{i}\Phi_{ij}^{(1)}}+\mathrm{e}^{-\mathrm{i}\Phi_{ij}^{(2)}}}}.
\end{equation}

Ideally the two dressing pulses are identical, such that $\varphi^{(1)} = \varphi^{(2)} = \varphi$ and $\Phi_{jk}^{(1)} = \Phi_{jk}^{(2)} \equiv \Phi_{jk}$. Using the experimentally determined laser pulses we have studied effects of occurring field fluctuations and found that they have a negligible effect for all of the relevant observables. Hence, we neglect potential asymmetries between the two pulses such that Eq.~(\ref{EqEchoSignalCoherent}) reduces to
\begin{equation}\label{EqEchoCoherent}
\langle\hat{\sigma}_{\downarrow\downarrow}^{(i)}\rangle = \frac{1}{2} + \frac{1}{2}\prod_{j\neq i}\cos (\Phi_{ij}).
\end{equation}
Following the same procedure one can readily obtain the corresponding correlation function
\begin{widetext}
\begin{eqnarray}
g_{ij}^{(2)}&=& \langle \hat{\sigma}_{\downarrow\downarrow}^{(i)} \hat{\sigma}_{\downarrow\downarrow}^{(j)}\rangle - \langle \hat{\sigma}_{\downarrow\downarrow}^{(i)}\rangle \langle \hat{\sigma}_{\downarrow\downarrow}^{(j)}\rangle \nonumber \\
 &=& \frac{1}{8}\left(\prod_{k\neq i,j}\cos\Phi_{k,ij}^{(+)}+\prod_{k\neq i,j}\cos\Phi_{k,ij}^{(-)}\right)-\frac{1}{4}\cos\Phi_{ij}^2\prod_{k\neq i,j} \cos\Phi_{ik} \cos\Phi_{jk}\label{Eqg2Coherent}\\
&\approx&\frac{\Phi_{ij}^2}{4}\:\:({ \Phi_{ij}\ll1}),
\end{eqnarray}
\end{widetext}
where $\Phi_{k,ij}^{(\pm)}=\Phi_{ik}\pm\Phi_{jk}$, and $\Phi_{ii} = 0$. In the last step we have assumed $\Phi_{ij}\ll1$, valid for short dressing times, in which case the correlation function directly reflects the shape of the interaction potential.

\subsection{Including spontaneous emission}\label{SecSpontEm}

In the presence of dissipative effects, the induced spin dynamics still permits an analytical treatment \cite{Foss-Feig2013}. Here, we first focus on effects of spontaneous emission, i.e. spontaneous decay of a Rydberg state to one of the atomic ground states. In the following, it is important to distinguish between the bare (undressed) ground state $\ket{g_\uparrow} = \ket{F = 2, m_F = -2}$ and (dressed) eigenstate in the presence of the excitation laser light, $\ket{\uparrow}$. As illustrated in Fig.~\ref{fig:S8}a, we need to distinguish three different decay channels. This includes the decay of the Rydberg state to the targeted ground state $\ket{g_\uparrow}$ with a rate $\Gamma_\uparrow$, described by the jump operator 
\begin{align}
\C_{2} &= \sqrt{\Gamma_\uparrow} \ket{g_\uparrow}\bra{\Rydblvl}\;,
\end{align}
the decay of the Rydberg state to the uncoupled ground state $\ket{\downarrow}$ with a rate $\Gamma_\downarrow$, described by the jump operator 
\begin{align}
\C_{1} &= \sqrt{\Gamma_\downarrow} \ket{\downarrow}\bra{\Rydblvl}\;,
\end{align}
and the decay of the Rydberg state to the remaining ground state manifold denoted by $\ket{g_0}$ with a rate $\Gamma_0$, described by the jump operator 
\begin{align}
\C_{0} &= \sqrt{\Gamma_0} \ket{g_0}\bra{\Rydblvl}\;.
\end{align}
The corresponding master equation describing the time evolution of the density matrix during the dressing stage thus reads
\begin{widetext}
\begin{align}
\pdd{\hat{\rho}(t)}{t} = -\frac{\mathrm{i}}{\hbar} \comm{\hat{H}_{\rm dr}}{\hat{\rho}} + \sum_{i=1}^N\sum_{k} \br{\C_{k,i} \hat{\rho} \Cd_{k,i}  - \frac{1}{2}\br{\Cd_{k,i} \C_{k,i} \hat{\rho} + \hat{\rho} \Cd_{k,i}\C_{k,i}}}.\label{EqLindbladDiagonal}
\end{align}
\end{widetext}
Next we need to express the Lindblad operator in the basis of the dressed spin states. As illustrated in Fig.~\ref{fig:S8}b, this leads to an effective dephasing of the dressed $\ket{\uparrow}$ state and opens up additional decay channels to the remaining ground states. The corresponding jump operators are given by
\begin{align}
\J_{02} &= \sqrt{\Gamma_0}{\frac{\Omega}{2|\Delta|}} \ket{g_0}\bra{\uparrow},\label{EqJ02}\\
\J_{12} &= \sqrt{\Gamma_{\downarrow}}{\frac{\Omega}{2|\Delta|}} \ket{\downarrow}\bra{\uparrow},\\
\J_{22} &= \sqrt{\Gamma_{\uparrow}}{\frac{\Omega}{2|\Delta|}} \ket{\uparrow}\bra{\uparrow}.
\end{align}
In addition, one obtains a dissipative transition from the dressed ground state to the dressed Rydberg state $\ket{\tilde e}$, which is predominantly composed of $\ket{\Rydblvl}$. The associated jump operator is given by
\begin{align}
\J_{32} &= \sqrt{\gamma_{\uparrow}}\br{\frac{\Omega}{2|\Delta|}}^2 \ket{\tilde{e}}\bra{\uparrow},\label{EqJ32}
\end{align}
The decay processes out of $\ket{\tilde e}$ are approximately equal to the bare atomic jump operators 
\begin{eqnarray}\label{EqJk3}
\J_{k0} &=& \sqrt{\Gamma_{0}} \ket{g_0}\bra{\tilde e},\nonumber\\
\J_{k1} &=& \sqrt{\Gamma_{\downarrow}} \ket{\downarrow}\bra{\tilde e},\nonumber\\
\J_{k2} &=& \sqrt{\Gamma_{\uparrow}} \ket{\uparrow}\bra{\tilde e},
\end{eqnarray}
with the exception of an arising dephasing term
\begin{align}\label{EqJ33}
\J_{33} &= \sqrt{\Gamma_{\uparrow}}\br{\frac{\Omega}{2|\Delta|}} \ket{\tilde e}\bra{\tilde e}.
\end{align}
Since the adiabatic laser-coupling dressing does not populate the dressed Rydberg state and the dissipative coupling Eq.~(\ref{EqJ32}) is suppressed by a factor $\Omega/2|\Delta|$ we can neglect all jump operators involving $\ket{\tilde e}$. With this simplification the master equation in the dressed-state representation reads
\begin{widetext}
\begin{align}
\pdd{\hat{\rho}(t)}{t} = -\frac{\mathrm{i}}{\hbar} \comm{\hat{H}_{\rm dr}}{\hat{\rho}} + \sum_{m = 0}^2 \sum_{i} \gamma_m \br{\J_{m, i} \hat{\rho} \Jd_{m, i} - \frac{1}{2}\br{\Jd_{m,i} \J_{m,i} \hat{\rho} + \hat{\rho} \Jd_{m,i}\J_{m,i}}}, \label{EqLindblad}
\end{align}
\end{widetext}
where we have defined the operators for particle $i$
\begin{align}
\J_{0,i} &= \ket{g_0}_i\bra{\uparrow},\label{EqJ0}\\
\J_{1,i} &= \ket{\downarrow}_i\bra{\uparrow},\\
\J_{2,i} &= \ket{\uparrow}_i\bra{\uparrow},
\end{align}
and the effective rates
\begin{align}
\gamma_0 &= {\Gamma_{0}}\br{\frac{\Omega}{2\Delta}}^2,\label{EqRate0}\\
\gamma_\downarrow &= {\Gamma_{\downarrow}}\br{\frac{\Omega}{2\Delta}}^2,\\
\gamma_\uparrow &= {\Gamma_{\uparrow}}\br{\frac{\Omega}{2\Delta}}^2\label{EqRate2}.
\end{align}
This diagonal form of the Lindblad equation (\ref{EqLindblad}) enables the analytic treatment of the correlated spin dynamics. Using a Monte Carlo wave function approach \cite{Molmer1993a}, the expectation value of an operator is evaluated as a sum of coherent quantum trajectories interspersed with discrete jumps. Each trajectory is weighted by the corresponding jump probability, determined by the rates (\ref{EqRate0}) - (\ref{EqRate2}). Formally, for some operator $\hat{O}$,
\begin{equation}\label{EqQMC}
\expval{\hat{O}} = \sum_{\textrm{traj.}} P(\textrm{traj.}) \expval{\hat{O}}_{\textrm{traj.}},
\end{equation}
where $P(\textrm{traj.})$ is the probability for a given quantum trajectory and $\langle \hat{O} \rangle_{\textrm{traj.}}$ is the expectation value of $\hat{O}$ at the end of the trajectory. For our particular pulse sequences the expectation values of the individual trajectories can be calculated by applying all jump operators at the initial time \cite{Foss-Feig2013} . Each trajectory expectation value can then be evaluated using similar techniques to those  described for the unitary evolution, allowing for an analytical evaluation of the sum in Eq. (\ref{EqQMC}). 

\begin{figure}
\begin{center}
\includegraphics[width=0.45\textwidth]{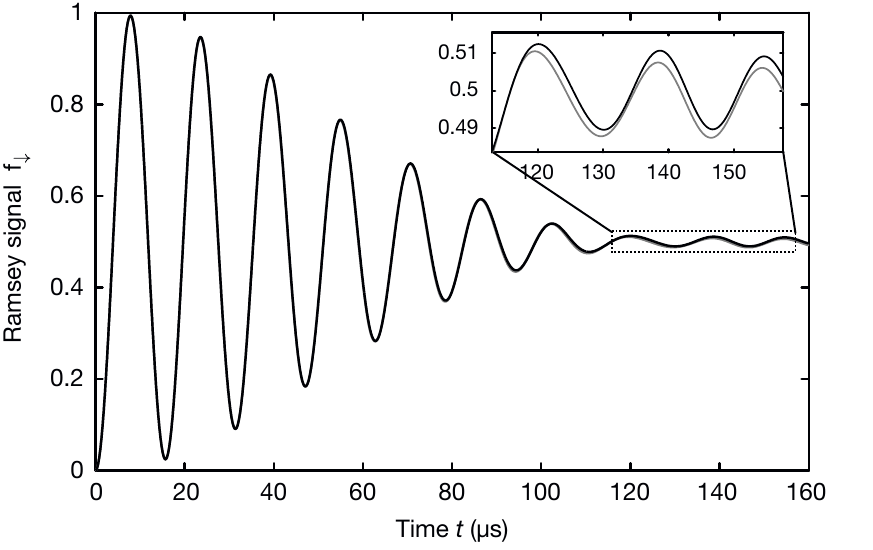}
\caption{\textbf{Influence of spontaneous decay on spin coherence.}
Ramsey signal, Eq.(\ref{eq:fdown}) as a function of the dressing time $t$. The black solid line shows the fully coherent result Eq.(\ref{EqRamseyCoherent}) and gray line the result in the presence of spontaneous emission, Eq.(\ref{EqRamseySpontEmission}). Calculations where performed for $31P_{1/2}$ Rydberg states using the effective spin-spin interaction potential shown in Fig.~\ref{fig:S4}a. For these and all other parameters of our experiment, effects of spontaneous emission are practically negligible, even at large times.} \label{fig:S9}
\end{center}
\end{figure}

Assuming that the dressing field is a simple square-pulse of duration $t$, the Ramsey signal in the presence of spontaneous emission can be written in a compact form 
\begin{align}\label{EqRamseySpontEmission}
\langle \hat{\sigma}_{\downarrow\downarrow}^{(i)} \rangle &= \frac{1}{4} + \frac{\ee^{-(\gamma_0 + \gamma_\downarrow) t}\gamma_0 + \gamma_\downarrow}{4 (\gamma_0 + \gamma_\downarrow)} - \frac{1}{2}\ee^{-(\gamma_0 + \gamma_\downarrow + \gamma_\uparrow) t/2}\  \mathrm{Re} \br{\Xi},
\end{align}
with
\begin{widetext}
\begin{equation}
\Xi = \frac{\mathrm{e}^{-\mathrm{i}\varphi }}{2^N} \prod_{j \neq{k}} \br{1 + \mathrm{e}^{-(\gamma_0 + \gamma_{\downarrow})t} \ee^{-\mathrm{i}\Phi_{jk} } + (\gamma_0 + \gamma_\downarrow)\frac{1 - \mathrm{e}^{-(\gamma_0+\gamma_{\downarrow})t}\ee^{-\mathrm{i}\Phi_{jk} }}{\gamma_0 + \gamma_\downarrow + \mathrm{i} \Phi_{jk} }}.
\end{equation}
\end{widetext}
In Fig.~\ref{fig:S9} we show the fraction
\begin{equation}\label{eq:fdown}
f_\downarrow=N^{-1}\sum_i\langle \hat{\sigma}_{\downarrow\downarrow}^{(i)} \rangle
\end{equation}
obtained from Eq.(\ref{EqRamseySpontEmission}) in comparison to the coherent result Eq. (\ref{EqRamseyCoherent}) for typical parameters of our experiment. Clearly, the influence of spontaneous emission is negligible for timescale considered in our measurements, highlighting the promise of Rydberg-dressing for observing unitary dynamics of interacting spins.

\subsection{Effects of black-body radiation}\label{SecBBrad}

As discussed in the main text, our experiments suggest the presence of additional dissipation that induces loss of a sizeable fraction of the atoms  but is triggered with a small probability. We attribute this loss to black-body radiation (BBR), which stimulates incoherent transitions to other Rydberg states. This process has fundamentally different consequences for Rydberg-dressing. As described in the previous section, spontaneous emission from the virtually excited Rydberg states causes comparably weak decoherence, in particular because the dominant quantum jump processes leave the atoms within the Rydberg-dressed ground state manifold.

The situation is profoundly different for BBR-induced transitions between Rydberg states. Here, the associated jump process fully projects the atomic state onto a Rydberg state, and thereby creates a real Rydberg atom out of a virtual Rydberg excitation. Not only does this take an atom out of the dressed ground state manifold, but also generates Rydberg states ($n^\prime S$ or $n^\prime D$) that feature direct dipole-dipole interactions with the laser-coupled $nP$ states. Such strong interactions are expected to cause substantial line broadening facilitating near-resonant laser-excitation of Rydberg states which eventually leads to rapid atom loss through decay out of the spin manifold ($\ket{\downarrow}$ and $\ket{\uparrow}$) and mechanical motion induced by the optical lattice potential or the Rydberg-Rydberg interactions.

Here we do not attempt to describe the detailed dynamics of these loss processes, but propose a phenomenological model assuming an instantaneous loss of all atoms in the Rydberg-dressed state $\ket{\uparrow}$, triggered by one-body BBR-induced transitions with a rate $\gamma_{\rm BB}$. 
Within the above Monte Carlo wave function approach \cite{Molmer1993a} this process is described by a projection of a given $N$-body state $\ket{\Psi}$ onto $\br{\ket{\downarrow}\bra{\downarrow} + \ket{0}\bra{\uparrow}}^N \ket{\Psi}$. The complete loss of all population in the $\ket{\uparrow}$ state stops any further dynamics during the dressing stage, which greatly reduces the complexity of the problem and the number of possible quantum trajectories. 

For both pulse sequences used in our experiments (Fig.~\ref{fig:S1}) each atom carries a $50\%$ $\ket{\uparrow}$-population during the dressing stage. The probability for a BBR transition to occur is thus
\begin{equation}
p_{\rm BB}=\frac{1}{2}-\frac{1}{2}\exp\!\left(-\gamma_{\rm BB}\int_0^{t/2} \beta(\tilde{t})^2 {\rm d}\tilde{t}\right)\;,
\end{equation}
where $\beta(t)^2=\Omega(t)^2/(4\Delta^2)$ is the Rydberg-state population of the dressed $\ket{\uparrow}$-state defined in the main text and the integration extends over a single dressing pulse of duration $t/2$. Since the decay can occur only once per dressing pulse, the probability that \emph{none} of the $N$ atoms undergoes a BBR-induced transition during a single dressing period is 
\begin{equation}
P_0 = (1-p_{\rm BB})^N \approx \exp\br{-\frac{N}{2} \gamma_{\rm BB}\int_0^{t/2} \beta(\tilde{t})^2 {\rm d}\tilde{t}}.
\end{equation}

\subsubsection{Ramsey spectroscopy}\label{SecRamseyBB}

For the Ramsey sequence (Fig.~\ref{fig:S1}a) the collective loss leaves an average of $N/2$ atoms in the $\ket{\downarrow}$-state, half of which is subsequently transferred to the $\ket{\uparrow}$-state by the final $\pi/2$ pulse, giving a final population of $\langle\hat{\sigma}_{\downarrow\downarrow}^{(i)}\rangle=1/4$. The total Ramsey signal is simply obtained, according to Eq. (\ref{EqQMC}), as the weighted sum of the signals with and without a BBR-transition
\begin{equation}\label{EqRamseyBB}
\langle\hat{\sigma}_{\downarrow\downarrow}^{(i)}\rangle = P_0 \langle\hat{\sigma}_{\downarrow\downarrow}^{(i)}\rangle_{\rm c} + \frac{1}{4}(1 - P_0),
\end{equation}
where $\langle\hat{\sigma}_{\downarrow\downarrow}^{(i)}\rangle_{\rm c}$ denotes the result of the unitary evolution given by Eq. (\ref{EqRamseyCoherent}). This expression has been used to obtain the theory data shown in Fig. 1d of the main text.

\subsubsection{Spin echo sequence}\label{SecEchoBB}
For the spin echo sequence we need to consider four different types of quantum trajectories: the completely unitary evolution, a BBR-transition occurring during the first dressing pulse, a BBR-transition occurring during the second dressing pulse and a BBR-transition occurring during both dressing pulses. As for the Ramsey sequence the unitary evolution yields a $50\%$ population in the Rydberg-dressed state such that the probability for a BBR-transition to occur during one of the two dressing pulses is given by $2P_0(1-P_0)$. Following the same arguments as above the final spin echo signal conditioned on having exactly one loss event is again $\langle\hat{\sigma}_{\downarrow\downarrow}^{(i)}\rangle=1/4$. Since the loss destroys all spin-correlations the corresponding two-body expectation values are simply $\langle\hat{\sigma}_{\downarrow\downarrow}^{(i)}\hat{\sigma}_{\downarrow\downarrow}^{(j)}\rangle=1/16$. If a BBR-transition occurs during both of the dressing-pulses it causes complete atom loss and, thus, does not contribute to any relevant observable. The remaining number of atoms after the two dressing pulses of total time $t$ is thus given by
\begin{equation}\label{EqEchoBB_N}
\begin{split}
N   &= P_0^2 N(0) + N(0) P_0 [1 - P_0]\\
	&= N(0) \exp\br{-\frac{N(0)}{2} \gamma_{\rm BB}\int_0^{t/2} \beta(\tilde{t})^2 {\rm d}\tilde{t}}\;,
\end{split}
\end{equation}
and the spin echo signal can be written as
\begin{equation}\label{EqEchoBB_spin}
\langle\hat{\sigma}_{\downarrow\downarrow}^{(i)}\rangle = P_0^2 \langle\hat{\sigma}_{\downarrow\downarrow}^{(i)}\rangle_{\rm c} + \frac{1}{2} P_0 [1 - P_0],
\end{equation}
where $\langle\hat{\sigma}_{\downarrow\downarrow}^{(i)}\rangle_{\rm c}$ is given by Eq. (\ref{EqEchoCoherent}). Eqs.~(\ref{EqEchoBB_N}) and (\ref{EqEchoBB_spin}) have been used for the theory curves in Fig. 4 of the main text.
Similarly we obtain for the spin correlation function
\begin{equation}
g_{ij}^{(2)} = P_0^2\left( g_{ij,{\rm c}}^{(2)}+\langle\hat{\sigma}_{\downarrow\downarrow}^{(i)}\rangle_{\rm{c}}\langle\hat{\sigma}_{\downarrow\downarrow}^{(j)}\rangle_{\rm c}\right) + \frac{1}{8}P_0 [1 - P_0] - \expval{\hat{\sigma}_{\downarrow\downarrow}^{(i)}}\expval{\hat{\sigma}_{\downarrow\downarrow}^{(i)}}, \label{Eqg2BB}
\end{equation}
where $\langle\hat{\sigma}_{\downarrow\downarrow}^{(i)}\rangle_{\rm c}$, $g_{ij, c}^{(2)}$ and $\langle\hat{\sigma}_{\downarrow\downarrow}^{(i)}\rangle$ are given by Eq.~(\ref{EqEchoCoherent}), Eq.~(\ref{Eqg2Coherent}) and Eq.~(\ref{EqEchoBB_spin}), respectively. Eq.~(\ref{Eqg2BB}) is used for the theoretical data shown Figs. 2 and 3a of the main text.

\end{document}